\documentclass[structabstract]{aa}  %2columns version

\usepackage{natbib}  % bibliography
\bibpunct{(}{)}{;}{a}{}{,} %  A&A style for bibliography
\usepackage{graphicx}   % figures
\usepackage{txfonts}  % roman font
\usepackage{lscape}    % table
\usepackage{longtable}   %table
\renewcommand\footnoterule{  \kern -3pt  \hrule \kern 2pt }

\newcommand{\Teff} {T$_{\rm eff}$\ }  	
         	
\newcommand{\dsol} {d$_{\odot}$}         	
\newcommand{\Rsol} {R$_{\odot}$}         	
\newcommand{\A} {\AA{}}                  	
         	
\newcommand{\HII} {H\,\textsc{ii}}        	
\newcommand{\Ha} {H$\alpha$}      		
\newcommand{\Hb} {H$\beta$}      		
\newcommand{\Hg} {H$\gamma$}      		
\newcommand{\Hd} {H$\delta$}      		
\newcommand{\HeI} {He\,\textsc{i}}		
\newcommand{\HeII} {He\,\textsc{ii}}		
		
\newcommand{\NII} {[N\,\textsc{ii}]}		
\newcommand{\SII} {[S\,\textsc{ii}]}		
\newcommand{\SIII} {[S\,\textsc{iii}]}		
\newcommand{\SIV} {[S\,\textsc{iv}]}		
\newcommand{\OI} {[O\,\textsc{i}]}		
\newcommand{\OII} {[O\,\textsc{ii}]}		
\newcommand{\OIII} {[O\,\textsc{iii}]}

\newcommand{\NeIII} {[Ne\,\textsc{iii}]}	
		
\newcommand{\ArIII} {[Ar\,\textsc{iii}]}	
\newcommand{\ArIV} {[Ar\,\textsc{iv}]}	
\newcommand{\FeIII} {[Fe\,\textsc{iii}]}	
\newcommand{\ClIII} {[Cl\,\textsc{iii}]}	
\newcommand{\SiII} {[Si\,\textsc{ii}]}		
\newcommand{\CII} {C\,\textsc{ii}}

\begin{document}

\title{Chemical distribution of HII regions towards the Galactic anticentre}

%\subtitle{}
\author{A. Fern\'andez-Mart\'in \inst{1,2} \thanks{e-mail: alba@iaa.es} ,
							E. P\'erez-Montero  \inst{1},							
							J.M. V\'ilchez \inst{1} , 
							\and 
							A. Mampaso \inst{3,4}.
                			}
\institute{Instituto de Astrof\'isica de Andaluc\'ia (IAA-CSIC),  Glorieta de la Astronom\'ia  S/N, 18008 Granada, Spain
	    \and Instituto de Radioastronom\'ia y Astrof\'isica (IRyA-UNAM), 58089 Morelia, Mexico 
	    \and Instituto de Astrof\'sica de Canarias (IAC), 38200 La Laguna, Tenerife, Spain
	    \and Departamento de Astrof\'isica, Universidad de La Laguna, 38206 La Laguna, Tenerife, Spain
				}                                                                       
%email: alba@iaa.es

% \date{Received.....; accepted..... } %set by editors and publisher

% 5 {} token are mandatory

\abstract   
% Context
{The study of the radial variations of metallicity across the Galactic disc is a powerful method for understanding the history of star formation and chemical evolution 
of the Milky Way. Although several studies about gradients have been performed so far, the knowledge of the Galactic antincentre is still poor.} 
%Aims 
{This work aims to determine accurately the physical and chemical properties of a sample of \HII~regions located at R$_G>$11~kpc and to study the radial distribution of 
abundances in the outermost part of the Galaxy disc.}
%Methods
{We carried out new optical spectroscopic observations of nine \HII~regions with the William Herschel Telescope covering the spectral range from 3500~\A~to 10100\A. 
In addition, we increased the sample by searching the literature for optical observations of regions towards the Galactic anticentre, re-analysing them to obtain a 
single sample of 23 objects to be processed in a homogeneous and consistent manner. The total sample distribution covers the Galactocentric radius from 11~kpc to 18~kpc.}
%And Results
{Emission line ratios were used to determine accurate electron densities and temperatures of several ionic species in 13 \HII~regions. These physical parameters were 
applied to the spectra to determine direct total chemical abundances. For those regions without direct estimations of temperature, chemical abundances were derived by 
performing tailor-made photoionisation models and/or by using an empirical relation obtained from radio recombination and optical temperatures. 

We performed weighted least-squares fits to the distribution of the derived abundances along the Galactocentric distances to study the radial gradients of metallicity 
across the outermost part of the MW. The distributions O/H, N/H, S/H, and Ar/H towards the anticentre can be represented by decreasing linear radial gradients, while in 
the case of N/O abundances the radial distribution is better fitted with a two-zone model. The He/H radial gradient is presented here for the first time; we find a slope 
that is not significantly different from zero. The derived gradient for oxygen shows a clear decrease with distance with a slope of -0.053$\pm$0.009 dex~kpc$^{-1}$. Although 
a shallower slope at large Galactocentric distances is suggested by our data, the flattening of the distribution cannot be confirmed and more objects towards the anticentre 
need to be studied in order to establish the true form of the metallicity gradient.}
%(Conclusions)
{}


\titlerunning{Anticentre chemical distribution}  
\authorrunning{Fern\'andez$-$Mart\'in et al.}  
\maketitle  

%
%________________________________________________________________

%SECTION 1
%%%%%%%%%%%%%%%%%%%%%%%%%%%%%%%%%%%%%%%%%%%%%%%%%%

\section{Introduction} \label{sect:intro} 
The chemical evolution of the interstellar medium (ISM) varies between galaxies and is both position and time dependent within a galaxy. Since the nucleosynthesis of different 
elements occurs in stars of different masses, the study of chemical abundances in the discs of spiral galaxies is a powerful method for understanding the history of star formation 
and evolution of galaxies.

A knowledge of the radial variations of metallicity across the galactic disc (i.e. abundance gradients) is central to our understanding of a wide variety of observed phenomena, 
including the  physics of star formation rates \citep{Phillipps1991},  initial mass function \citep{Guesten1982}, the radial inflows/outflows of gas \citep{Mayor1981}, and the 
stellar evolution and the process of nucleosythesis itself. Galactic abundance gradients in the ISM were first described by \citet{Searle1971} in a survey of \HII~regions 
in six late-type spiral galaxies. Since then, considerable effort has been made to establish the chemical distribution in the Milky Way (MW) by studying many sources such as 
supernova remnants (SNR) \citep{Binette1982}, molecular clouds \citep{Rudolph1996}, open clusters \citep{Twarog1997, Andreuzzi2011}, cepheids \citep{Luck2003,Korotin2014}, 
OB stars \citep{Rolleston2000}, and planetary nebulae (PNe) \citep{Maciel1994,Henry2010}. However, when deriving the abundances representative of the current ISM values, 
\HII~regions provide the most accessible probe of abundances gradients. Since they are bright and hot they emit strongly in many lines observable over much of the MW. 
Unlike stars, \HII~regions probe the current state of abundances, and unlike PNe and SNR, do not contaminate the surrounding ISM.\\

The existence of a large-scale gradient in the MW with \HII~regions was established by the pioneering work of \citet{Shaver1983}, who found a decrease of metallicity 
with Galactic distances. Subsequently, several studies with \HII~regions \citep{Hawley1978,Talent1979,Fich1991,Hunter1992,Afflerbach1997,Vilchez1996,Deharveng2000,Rudolph2006} 
have been carried out, firmly establishing the existence of a negative gradient of abundance of the elements heavier than helium in the disc of the MW.

\begin{table*}[h!t!]
\caption{Names and positions of the \HII~regions observed with WHT-ISIS and log of observations.} 
\label{table:logISIS} 
\centering 
\begin{tabular}{l c c c c c }
\\
\hline
\HII~region$^{(a)}$   &   $\alpha_{J2000}$    &      $\delta_{J2000}$     & Airmass$^{(b)}$  & Exp. Time $^{(c)}$    & Date\\ 
                           &          (h~m~s)         &       ($^{\rm o~'~''}$)   &                      &       (s)                &          \\	
\hline\hline
\\
S83                     &   19:24:30.77            &   +20:47:45.92             &   1.18             &  3$\times$1800    			            					 & July, 13, 2010 \\	 
S132                   &   22:19:08.26            &   +56:05:12.98             &   1.37             &  3$\times$1800                 	             				& July, 13, 2010 \\
S156                   &   23:05:08.33            &   +60:14:46.29             &   1.23             &  3$\times$700 / 3$\times$200 $^{(d)}$         					& July, 13, 2010 \\
S162                   &   23:20:43.94            &   +61:12:27.06             &   1.20             & 500+900+1000 / 18$\times$100 $^{(d,e)}$           & July, 13, 2010 \\
S207                   &    4:19:49.49             &   +53:09:34.91             &   1.27            &  3$\times$1200                  	             				& Dec., 19, 2009 \\
S208                   &    4:19:32.39             &   +52:58:38.98             &   1.21            &  3$\times$1500                  	             				&  Dec., 19, 2009 \\
S212                   &    4:40:36.50             &   +50:27:44.31             &   1.14            &  3$\times$1500                  	             				&  Dec., 19, 2009 \\
S228                   &    5:13:23.37             &   +37:27:19.63             &   1.02            &  3$\times$1500                  	            				 &  Dec., 19, 2009 \\
S270                   &    6:10:12.96             &   +12:48:37.38             &   1.24            &  1200+1300+1500 $^{(e)}$                        	& Dec., 19, 2009 \\
\hline
\end{tabular}
\begin{list}{}{}\footnotesize{
\item $^{(a)}$ Sh2-\textit{number} is the IAU standard notation for objects in \citet{Sharpless1959} catalogue. For consistence with other works and simplification, in this 
paper we identify them as S-\textit{number}.
\item $^{(b)}$ Mean value during the observations.
\item $^{(c)}$ In all the regions, except S156 and S162, the blue and red arms were acquired simultaneously with the same exposition time. 
\item $^{(d)}$ Individual times for each grating are indicated for S156 and S162 separated by a bar (blue/red).
\item $^{(e)}$ Regions S162 and S270 were observed with different exposition times (with the same arm). This was taking into account when combining images in the data reduction.
}
\end{list}
\end{table*}

Nonetheless, the sampling of the whole Galactic disc is still poor. In particular, relatively few \HII~regions have been studied towards the Galactic anticentre region. 
Optically, there are only ten Galactic \HII~regions which have been observed at R$_G>$11 kpc and for which direct measurements of the electron temperature are 
available. This is also a handicap for the MW  studies, compared with those for nearby external galaxies, and limits the application of models of galactic evolution. In 
particular, it limits those aspect of the models that describe the evolution of the outermost parts of the disc, which might be considered  -- at least from the point of 
view of the chemical evolution -- to be much closer to the pre-galactic/early conditions in the MW.  Moreover, the possibility of variations in the slope of the gradients 
in the outer disc has been a subject of debate in the past years. Some authors have claimed that radial abundance gradients of some elements may flatten out at the outer 
parts of the Galactic disc \citep{Fich1991,Vilchez1996}, while other authors do not support such flattening \citep{Deharveng2000,Rudolph2006}.

Therefore, extending the measurement of abundances to large Galactocentric distances in the MW is essential to our understanding of the metallicity gradient and the chemical 
evolution of our Galaxy. \\

This paper has two main purposes. The first is to present new long-slit observations in a wide spectral range of \HII~regions located towards the Galactic anticentre with 
Galactocentric distances from 11~kpc to 17~kpc. The second is to present a re-determination of chemical abundances, from direct electron temperatures and from physical 
modelling, of all the optical outermost \HII~regions observations in the literature in a homogeneous and consistent manner to perform a self-consistent study of the chemical 
gradients towards the outermost disc of the MW. The wide optical spectral range covered, together with the consistent chemical analysis of the sample, allowed us to give an 
accurate description of the radial chemical distribution towards the Galactic anticentre up to a distance of 18~kpc.

In the following section we present the data sample including our own observations and other data from the literature. Section \ref{sect:results} describes the analysis and 
the results of the study, while Sect. \ref{sect:discussion} is devoted to discussing the resulting abundance gradients and their implications in the context of Galactic 
evolution. Finally, we describe the main conclusions of this work in Sect. \ref{sect:conclusions}.

\begin{figure*}  
\centering
\includegraphics[width=\textwidth]{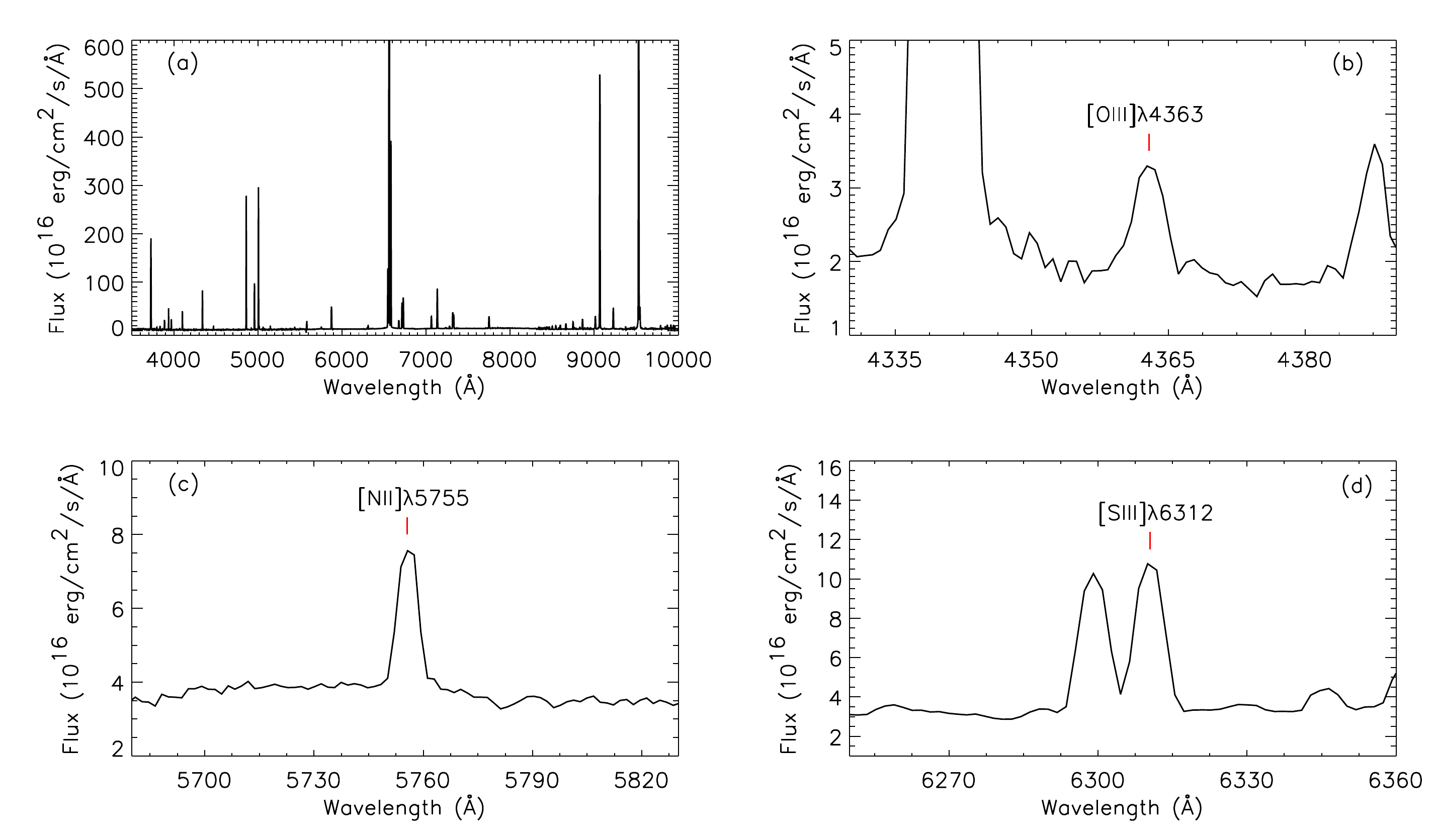}
\caption{Extracted observed flux-calibrated spectra. From left to right and top to bottom: (a) whole integrated spectrum of S156, (b) subset of the spectrum of S162 showing 
the temperature sensitive line \OIII$\lambda$4363, (c) subset of the spectrum of S162 showing the temperature sensitive line \NII$\lambda$5755, and (d) subset of the spectrum 
of S162 showing the temperature sensitive line \SIII$\lambda$6312.}
\label{fig:spectra}
\end{figure*}

%SECTION 2 
%%%%%%%%%%%%%%%%%%%%%%%%%%%%%%%%%%%%%%%%%%%%%%%%%%%%%%%%%%%%%%%%%%%%%%%%%%%%%%

\section{Sample of \HII~regions} \label{sect:sample} 
This work aims to study the chemical abundances of \HII~regions located towards the Galactic anticentre. Therefore we selected Galactic \HII~regions in the direction of 
the anticentre with R$_G>$ 11~kpc from catalogues of \citet{Fich1984}, \citet{Deharveng2000}, and \citet{Quireza2006}. The mother sample includes 89 anticentre \HII~regions 
observable from the northern hemisphere.

\subsection{Selection and observations of WHT data} \label{sect:owndata}
We selected for our observations those \HII~regions from the mother sample located at the furthest distance from the Galactic centre and, if possible, with available 
information on their thermal structure (e.g. a radio continuum temperature). Nine \HII~regions located at 11~kpc $ < $ R$_G< $ 17~kpc were observed; their names and 
positions are listed in Table \ref{table:logISIS}.

The observations were carried out in December 2009 and July 2010 using the ISIS double-armed spectrograph at the 4.2\,m William Herschel Telescope (WHT) at the Observatorio 
del Roque de los Muchachos (La Palma, Spain). A dichroic was set at $\sim$5336 \A~permitting simultaneous observation in the blue and red arms, which are optimised for 
their respective wavelength ranges. The blue arm detector, EEV12, is an array of 4096$\times$2048 (13.5 $\mu$m) pixels giving a spatial resolution of  0.2 arcsec/pix, while 
the CCD for the ISIS red arm, RED+, is a red-sensitive array of 4096$\times$2048 (15.0 $\mu$m)  with a spatial resolution of 0.22 arcsec/pix. The gratings were chosen in order 
to obtain as much information as possible on the most important emission lines of the optical range. In the blue arm, the R300B grating  was centred at 4400 \A~covering 
the effective spectral range from 3200 \A~to 5800 \A~with a dispersion of 0.86 \A/pix, giving a spectral resolution of $\mathrm{R=\lambda/\delta\lambda\sim}$1070 at 4400 \A. 
In the red arm, the R158R was centred at 7800 \A~covering the effective spectral range from 5300 \A~to 10000 \A~with a dispersion of 1.81 \A/pix and a spectral resolution 
of R$\sim$1010 at 7800 \A.

Table \ref{table:logISIS} also shows the observational log for the \HII~regions of the observed sample. At least three exposures were taken in each region with different 
exposition times depending on the object.  The slit width was set at 1 arcsec throughout the observation and oriented in parallactic angle  to avoid differential atmospheric 
refraction problems. The necessary bias frames, continuum, arcs, and spectrophotometric standard stars were also acquired.\\ 

The spectra were reduced using {\tiny IRAF}\footnote{The Image Reduction and Analysis Facility IRAF is distributed by the National Optical Astronomy Observatories, which 
are operated by the Association of Universities for Research in Astronomy, Inc., under cooperative agreement with the National Science Foundation.} by following the standard 
procedure for 2D spectroscopic observations (bias subtraction, flat-fielding, and cosmic ray rejection). The wavelength calibration was made using CuNe+CuAr arcs taken 
shortly after the object spectrum was taken. Two spectrophotometric standard stars were observed each night (G191-B2B and HR345 in December 2009, and BD+26\,2603 and 
BD+17\,4708 in July 2010) so that the spectra could be flux calibrated. 

For each exposure, a background region was selected in order to avoid nebular emission or stars. Pixels of these regions were combined with a median in a single spectrum and 
a 2D background spectrum was created with the value of the combined spectrum in each column. This sky  background was subsequently subtracted from every science object. 
This method was used in all the \HII~regions except for S162 where the S156-sky was used for the sky subtraction since S162 was too extended and no area without 
nebular emission was found. The sky subtraction worked well overall; only the strongest sky-lines in the far red spectra ($\lambda >$9000 \A) left some residuals. 

In order to select the zones to extract the 1D spectra we examined the spatial profiles of the most important emission lines, whenever possible selecting  
areas with emission of the auroral lines necessary to estimate the electron temperature. The spectra were traced and extracted using the {\tiny APALL} task in {\tiny IRAF} 
with an aperture optimised for a good S/N which was able to minimise the contamination of adjacent pixels; in this process we took into account the different spatial scales 
between blue and red arms. Eighteen spectra were finally extracted over the sample. In Fig. \ref{fig:spectra} we present four representative examples showing the auroral 
lines necessary to estimate 
electron temperatures.

\subsection{Additional data from the literature} \label{sect:extended}
To realise a more extensive study of the Galactic anticentre, the sample was increased with \HII~regions from previous works. With this aim, we carried out an exhaustive 
bibliographical review, selecting those \HII~regions located at R$_G>$11~kpc observed in the optical range and with measurements of the electron temperature sensitive 
lines or, at least, with information about the \SIII$\lambda$9068 line to obtain chemical abundances by means of individual photoionozation models. 

As a result of this search we found 14 \HII~regions that satisfy the stipulated requirements: S98, S127, S128, S209, S219, S266 (B), S283, and S203 (BFS31) from \citet{Vilchez1996}; 
S255 (c), S301 (RCW6), and S311 (RCW16a) from \citet{Shaver1983}; S158 (P1) from \citet{Talent1979}; S206 from \citet{Caplan2000}; and S298 (B2) from \citet{Esteban1990}. 
Some authors provide information about regions in several pointings; in these cases we chose those with more emission lines measured (in parentheses we indicate the 
pointing selected for our study and the identification given by each author).

\begin{table*}[ht!]
\caption{Fundamental physical parameter of the \HII~regions. Numbers indicate the corresponding references (see footnote and text for details).} 
\label{table:infoRHII} 
\centering 
\begin{tabular}{l  c c c c c c c c c}
\\
\hline
Region	&	l	            &	b$^{[1]}$	      &	R$_G$ 	   &	\dsol 	     &	  z$^{[7]}$	   &	$\Theta^{[1]}$	&	Spectral type	 of dominant    & t$_e^*$ &   Other \\
               &  ($^o$)    &    ($^o$)            &        (kpc)     &   (kpc)       &     (pc)              &   (arcsec)             &               exciting star                 &   ($\times$10$^4$K)       &          names       \\	
\hline\hline
\\
S83	&	55.12	&	2.42	&	15.2	$^{[2]}$	&	18.7	$^{[2]}$	&	-	&	2	&	-	&	-	&	-	\\
S98	&	68.15	&	1.02	&	12.8	$^{[3]}$	&	13.3	$^{[3]}$	&	-	&	15	&	O4V$^{[8]}$	&	1.08$^{[3]}$	&	-	\\
S127	&	96.29	&	2.60	&	13.9	$^{[2]}$	&	12.2	$^{[2]}$	&	698	&	2	&	O8V$^{[9]}$	&	1.14$^{[3]}$	&	-	\\
S128	&	97.50	&	3.16	&	12.7	$^{[2]}$	&	9.0	$^{[2]}$	&	484	&	1	&	O7V$^{[9]}$	&	1.04$^{[3]}$	&	-	\\
S132	&	102.79	&	-0.65	&	11.3	$^{[4]}$	&	5.8	$^{[4]}$	&	-59	&	90	&	O8.5V$^{[13]}$/B0III$^{[14]}$/WN6$^{[15]}$	&	-	&	-	\\
S156	&	110.11	&	0.06	&	11.5	$^{[4]}$	&	5.3	$^{[4]}$	&	6	&	2	&	O8V$^{[10]}$/O6.5V:$^{[15]}$/O7V$^{[16]}$	&	0.92$^{[3]}$/0.91$^{[4]}$	&	-	\\
S158	&	111.55	&	0.82	&	12.4	$^{[4]}$	&	6.4	$^{[4]}$	&	38	&	10	&	O9V$^{[10]}$	&	0.85$^{[3]}$/0.82$^{[4]}$	&	NGC7538	\\
S162	&	112.22	&	0.23	&	11.1	$^{[4]}$	&	4.7	$^{[4]}$	&	13	&	40	&	O7I$^{[9]}$/O6.5IIIf$^{[11,16]}$	&	0.86$^{[3]}$/0.81$^{[4]}$	&	NGC7635	\\
S203	&	143.81	&	-1.57	&	11.3	$^{[5]}$	&	3.3	$^{[5]}$	&	-	&	2	&	B2V$^{[10]}$	&	-	&	BFS31	\\
S206	&	150.58	&	-0.94	&	11.1	$^{[2]}$	&	2.8	$^{[2]}$	&	-44	&	50	&	O4$^{[15]}$/O5V$^{[16]}$/O5neb$^{[17]}$	&	1.00$^{[3]}$/0.97$^{[4]}$	&	-	\\
S207	&	151.19	&	2.13	&	16.8	$^{[2]}$	&	4.3	$^{[2]}$	&	-	&	4	&	O9V$^{[12,13]}$/O9.5IV$^{[9,17]}$	&	-	&	-	\\
S208	&	151.31	&	1.99	&	16.8	$^{[2]}$	&	4.1	$^{[2]}$	&	261	&	1	&	O9.5V$^{[9]}$/B0V$^{[12,17]}$	&	-	&	-	\\
S209	&	151.60	&	-0.25	&	16.9	$^{[4]}$	&	8.9	$^{[4]}$	&	-51	&	14	&	O9III$^{[9]}$	&	1.08$^{[3]}$/1.05$^{[4]}$	&	-	\\
S212	&	155.35	&	2.60	&	16.7	$^{[4]}$	&	8.6	$^{[4]}$	&	278	&	5	&	O6I$^{[9]}$/O7f$^{[15]}$/O5.5neb$^{[17]}$	&	1.03$^{[3]}$/1.05$^{[4]}$	&	-	\\
S219	&	159.36	&	2.60	&	13.3	$^{[2]}$	&	4.5	$^{[2]}$	&	189	&	3	&	B2.5V$^{[10]}$/B0V$^{[16,17]}$	&	-	&	-	\\
S228	&	169.21	&	-0.90	&	13.8	$^{[4]}$	&	5.3	$^{[4]}$	&	-	&	8	&	O8Ve$^{[9,13]}$/B0V$^{[15]}$	&	0.94$^{[3]}$/0.97$^{[4]}$	&	-	\\
S255	&	192.64	&	0.00	&	11.0	$^{[5]}$	&	2.5	$^{[5]}$	&	-2	&	3	&	B0V$^{[15]}$/B0IIIneb$^{[17]}$	&	-	&	-	\\
S266	&	195.66	&	-0.09	&	17.9	$^{[6]}$	&	9.6	$^{[6]}$	&	-15	&	1	&	BeI? $^{[18]}$	&	-	&	-	\\
S270	&	196.83	&	-3.10	&	14.2	$^{[6]}$	&	5.9	$^{[6]}$	&	-478	&	1	&	-	&	-	&	-	\\
S283	&	210.81	&	-2.56	&	17.0	$^{[6]}$	&	9.1	$^{[6]}$	&	-407	&	3	&	B3V$^{[10]}$/O7V$^{[10]}$/B1V$^{[10]}$/B0:V:$^{[17]}$	&	-	&	-	\\
S298	&	227.75	&	-0.15	&	12.4	$^{[4]}$	&	5.0	$^{[4]}$	&	-7	&	22	&	WN5$^{[14,17]}$	&	1.25$^{[3]}$/1.36$^{[4]}$	&	NGC2359/RCW5	\\
S301	&	231.45	&	-4.41	&	12.9	$^{[4]}$	&	5.8	$^{[4]}$	&	-439	&	9	&	O6V$^{[16]}$/O7$^{[17]}$/B1III$^{[17]}$	&	0.91$^{[3]}$/0.97$^{[4]}$	&	RCW6	\\
S311	&	243.16	&	0.36	&	12.0	$^{[4]}$	&	5.4	$^{[4]}$	&	32	&	45	&	O5V$^{[13]}$/O6.5V$^{[13]}$/O6:$^{[14]}$	&	1.05$^{[3]}$/1.02$^{[4]}$	&	RCW16	\\
\hline
\end{tabular}
\begin{list}{}{}\footnotesize{ 
\item References: [1]=\citet{Blitz1982}, [2]=\citet{Caplan2000}, [3]=\citet{Balser2011}, [4]=\citet{Quireza2006}, [5]=\citet{Rudolph2006}, 
[6]=\citet{Fich1991}, [7]=\citet{Fich1984}, [8]=\citet{Mampaso1991}, [9]=\citet{Chini1984}, [10]=\citet{Russeil2007}, [11]=\citet{Conti1974}, [12]=\citet{Crampton1978}, 
[13]=\citet{Crampton1974}, [14]=\citet{Georgelin1970}, [15]=\citet{Hunter1990}, [16]=\citet{Georgelin1975}, [17]=\citet{Moffat1979}, [18]=\citet{Vilchez1996}.}
\end{list}
\end{table*}

\subsection{Final sample} \label{sect:totalsample}
The final sample to be analysed in this work includes both samples described above (Sects. \ref{sect:owndata} and \ref{sect:extended}) and consists of 23 \HII~regions 
located in a range of Galactoncentric distances from 11 kpc to 18 kpc. In Table \ref{table:infoRHII} we present a summary of the fundamental physical parameters of the 
\HII~regions analysed in this study. Each \HII~region is identified in Col. 1 by its Sharpless number \citep{Sharpless1959}. Columns 2 and 3 give the Galactic 
coordinates (l,b). Columns 4 and 5 give the Galactocentric and heliocentric distances (R$_G$ and \dsol), both of which are analysed and justified in Sect. \ref{sect:distances}. 
Column 6 shows the distance above the Galactic plane (z), and Col. 7 gives the angular size of the nebula ($\Theta$). In Col. 9 we show the spectral type of the dominant 
exciting star. Column 10 gives the electron temperature derived from radio observations (t$_e^*$) and, finally, Col. 11 indicates other identification names. The 
corresponding references are given in the footnotes.\\

%SECTION 3
%%%%%%%%%%%%%%%%%%%%%%%%%%%%%%%%%%%%%%%%%%%%%%%%%%%%%%%%%%%%%%%%%%%%%%%%%%%%%%

\section{Analysis and results} \label{sect:results}

\setcounter{table}{3}
\begin{table*}[H!t!]
\caption{Electron densities and electron temperatures.} 
\label{table:physical_parameters}
\centering                
\begin{tabular}{l c c c c c c}
\\
\hline
\HII~region &       n$_e$          &        t$_e$(\NII)                  &        t$_e$(\OIII)           &     t$_e$(\SIII)           &        t$_e$(\OII)              &      t$_e$(\SII)  \\
                &    (cm$^{-3}$)        &      (10$^4$~K)       &             (10$^4$~K)               &    (10$^4$~K)           & (10$^4$~K)                &                 (10$^4$~K)                \\
\hline \hline   \\
S83	&	256	$\pm$ 	62		&	1.16	$\pm$ 	0.03	$_E$	&	1.14	$\pm$ 	0.05		&	1.13	$\pm$ 	0.02		&	-				&	-			\\
S127	&	325	$\pm$ 	75		&	-				&	1.01	$\pm$ 	0.13	$_E$	&	0.88	$\pm$ 	0.15	$_E$	&	0.91	$\pm$ 	0.05		&	-			\\
S128	&	$<$100				&	-				&	0.92	$\pm$ 	0.15	$_E$	&	0.77	$\pm$ 	0.18	$_E$	&	0.98	$\pm$ 	0.12		&	-			\\
S132	&	315	$\pm$ 	15		&	0.96	$\pm$ 	0.05		&	0.82	$\pm$ 	0.07	$_E$	&	0.81	$\pm$ 	0.02		&	-				&	0.84	$\pm$ 	0.09	\\
S156	&	1133	$\pm$ 	29		&	0.96	$\pm$ 	0.04		&	0.92	$\pm$ 	0.03		&	0.80	$\pm$ 	0.01		&	-				&	0.87	$\pm$ 	0.18	\\
S158	&	1323	$\pm$ 	265		&	0.99	$\pm$ 	0.15		&	0.87	$\pm$ 	0.21	$_E$	&	0.72	$\pm$ 	0.25	$_E\dagger $	&	-				&	-			\\
S162	&	1440	$\pm$ 	23		&	0.93	$\pm$ 	0.03		&	0.94	$\pm$ 	0.02		&	0.80	$\pm$ 	0.03		&	-				&	0.81	$\pm$ 	0.03	\\
S206	&	357	$\pm$ 	161		&	-				&	0.95	$\pm$ 	0.02		&	-				&	0.87	$\pm$ 	0.05	$_E$	&	-			\\
S212	&	211	$\pm$ 	29		&	1.11	$\pm$ 	0.06		&	1.21	$\pm$ 	0.04		&	1.18	$\pm$ 	0.03		&	-				&	-			\\
S255	&	321				&	0.82				&	0.65			$_E$	&	-				&	-				&	-			\\
S298	&	$<$100				&	-				&	1.20	$\pm$ 	0.03		&	1.11	$\pm$ 	0.04	$_E$	&	1.17	$\pm$ 	0.08	$_E$	&	-			\\
S301	&	$<$100				&	-				&	$<$0.91				&	$<$0.77			$_E$	&	$<$0.96			$_E$	&	-			\\
S311	&	130				&	0.95				&	0.87				&	0.72			$_E$	&	-				&	0.88			\\
\\
\hline
\end{tabular} 
      \begin{list}{}{} \footnotesize{
                  \item {$_{E}$} Electron temperatures derived from other t$_e$ (see text for details).
                  \item {$\dagger$} In region S158  t$_e$(\SIII) has S/N$<$3. This fact was taken into account when deriving abundances.}
		\end{list}
		\end{table*}

\subsection{Line intensities and reddening correction} \label{sect:intensities} 
We now report on the method used to obtain the line intensities of the regions observed with ISIS (our own sample). Line fluxes were measured by integrating all the 
flux in the line between two given limits and over a fitted local continuum. Partially blended lines, such \Ha~and \NII$\lambda\lambda$6548,6584, were deblended with 
two or three Gaussian profiles to measure the individual line fluxes. All the measurement were made with the {\tiny SPLOT} task of {\tiny IRAF}. The statistical errors 
associated with the observed fluxes were calculated using the expression
\begin{equation}
\sigma_{\mathrm{1}}=\sigma_{\mathrm{c}} N^{1/2} [1+EW/(N\Delta)]^{1/2},
\end{equation}
where $\sigma_{\mathrm{1}}$ represents the error in the observed line flux, N is the number of pixels used to measure the line, EW the line equivalent width, 
$\sigma_{\mathrm{c}} $ the standard deviation of the continuum in a box near the line, and $\Delta$ represents the dispersion in \A/pix \citep{PerezMontero2003}.

The reddening coefficient c(\Hb) was derived from the observed flux ratios of the brightest Balmer lines \Ha/\Hb,  \Hg/\Hb, and  \Hd/\Hb~as compared with the 
theoretical values obtained from the public software of \citet{Storey1995} assuming Case B recombination and following an iterative process with the n$_e$ and t$_e$ 
estimations (Sect. \ref{sect:parameters}). To minimise the reddening effect and to avoid error propagation due to differences between red and blue grating calibrations, 
the measured lines were reddening corrected relative to \Hb, \Ha, or P10 (if S/N$>$5)  depending on the spectral range and later normalised to \Hb~flux using the 
corresponding theoretical ratios.

Table \ref{table:sample_intensities} lists the reddening-corrected intensities of the emission lines measured for every observed \HII~region labelled with their standard 
identification. The third column gives the adopted reddening function, f($\lambda$), using the extinction law by \citet{Cardelli1989} with $R_{\mathrm{V}}=3.1$. Errors 
in the emission line intensities were derived by propagating the observational errors in the fluxes and the reddening constant uncertainties. The estimated fluxes and 
errors were normalised to F(\Hb)=1000. The values obtained for c(\Hb) are also presented in the last row of Table \ref{table:sample_intensities}. 

At zero redshift the \SIII$\lambda$9530\A~line fluxes are strongly affected by a sky absorption band. Therefore, this line was not used for the derivation of the \SIII~electron 
temperature neither for the S$^{++}$/H$^+$ ionic abundance. For these quantities a theoretical ratio \SIII$\lambda$9530/\SIII$\lambda$9069=2.44 was assumed.\\

The methodology used to obtain the intensities of the \HII~regions selected from the literature was different since in these regions we did not measure the fluxes in spectra. 
In this case, we calculated the original reddened fluxes considering the extinction law and c(\Hb) used by each author and later we re-derived the intensities following 
the same self-consistent method as in our regions: theoretical Balmer lines with Case B recombination, iterative process with n$_e$ and t$_e$, and reddening function 
using the extinction law by \citet{Cardelli1989} with $R_{\mathrm{V}}=3.1$. Finally, we obtained consistent intensities derived exactly under the same conditions as in our 
sample. Data from these 14 \HII~regions were added to our new observations reported here to create  a data set of 23 \HII~regions with 11~kpc $< R_G<$ 18~kpc. Hereafter 
we work with all of these objects as a single sample following the same methodology for both samples.

\subsection{Physical parameters} \label{sect:parameters}
To obtain the physical conditions of the gas, we performed an iterative process for each region until an agreement was achieved between electron density (n$_e$) and electron 
temperature (t$_e$). The values for t$_e$ and n$_e$ derived for each \HII~region are shown in Table \ref{table:physical_parameters}.  Those regions without measurements 
of auroral lines necessary to estimate at least one electron temperature were discarded temporarily. Physical parameters and chemical abundances were not derived with 
this direct method for any of them and, therefore, they are not included in Table \ref{table:physical_parameters} (but see Sects. \ref{sect:models} and \ref{sect:tradio}).\\

Electron density, n$_e$, was calculated for all the observed regions from the \SII $\lambda\lambda$6717,6731 line ratio using the {\tiny IRAF} package {\tiny TEMDEN} based 
on a five-level statistical equilibrium model \citep{deRobertis1987,Shaw1995}. Electron temperatures, t$_e$, were derived using the appropriate line 
ratios $\mathrm{R_{N2}, R_{O3}, R_{S3}, R_{O2}}$, and $\mathrm{R_{S2}' }$ following the equations below for each ion: 
	\begin{equation}
	\label{eq:Te_conNIIbis}
	t_e(\NII)\,=\,0.537\,+\,0.000253~R_{N2}\,+\,42.126/R_{N2};
	\end{equation}
	\begin{equation}
	\label{eq:Te_conOIIIbis}
	t_e(\OIII)\,=\,0.8254\,-\,0.0002415~R_{O3}\,+\,47.77/R_{O3};
	\end{equation}
	\begin{equation}
	\label{eq:Te_conSIII}
	t_e(\SIII)\,=\left[R_{S3}\,+\,36.4\right]/\left[1.8~R_{S3}\,-\,3.01\right];
	\end{equation}
	\begin{eqnarray} 
	\label{eq:Te_conOII}
	t_e(\OII)\,=&\,a_0(n_e)\,+&\,a_1(n_e)~R_{O2}\,+\,a_2(n_e)/R_{O2},\nonumber\\ 
	\rm{where} & a_0(n_e) =& 0.23-0.005~n_e-0.17/n_e \nonumber\\
				&  a_1(n_e)=& 0.007+0.000009~n_e+0.0064/n_e \nonumber\\ 
				&  a_2(n_e)=& 38.3-0.021~n_e-1.64/n_e;
	\end{eqnarray}
	\begin{eqnarray}
 	\label{eq:Te_conSII}
	t_e(\SII)\,=&a_0(n_e)\,+&\,a_1(n_e)~R_{S2}'\,+\,a_2(n_e)/R_{S2}'\,+\,a_3(n_e)/R_{S2}'^2, \nonumber\\ 
	\rm{where} &  a_0(n_e)=&1.92-0.0017~n_e+0.848/n_e \nonumber\\ 
				& a_1(n_e)=&-0.0375+4.038~10^{-5}~n_e-0.0185/n_e \nonumber\\ 
				& a_2(n_e)=&-14.15+0.019~n_e-10.4/n_e \nonumber\\ 
				& a_3(n_e)=&105.64+0.019~n_e+58.52/n_e.
	\end{eqnarray}
In all the equations, electron density is in units of cm$^{-3}$ and electron temperatures in units of 10${^4}$~K.\\

\begin{figure*}
\centering
\includegraphics[width=6cm]{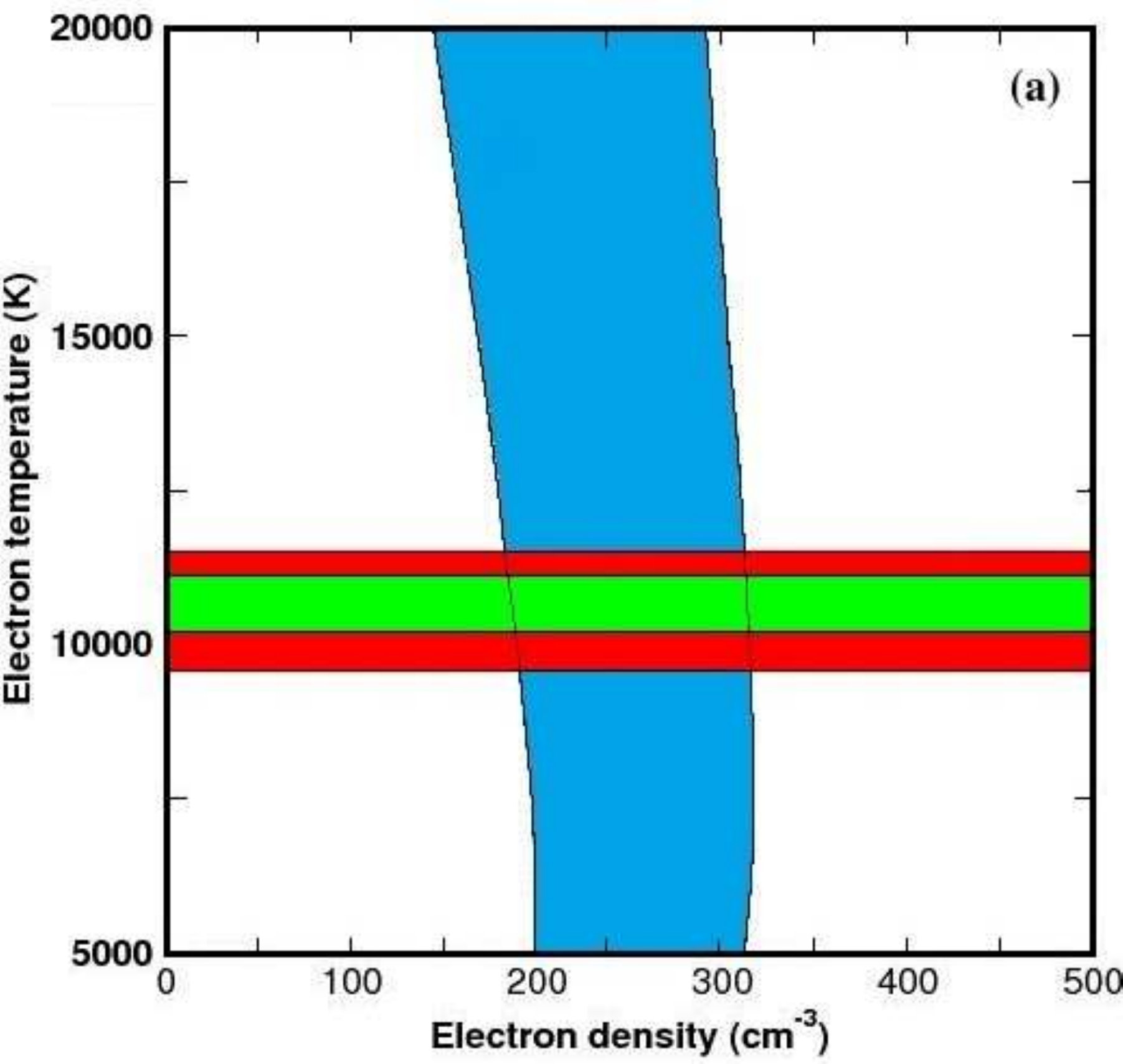}
\includegraphics[width=6cm]{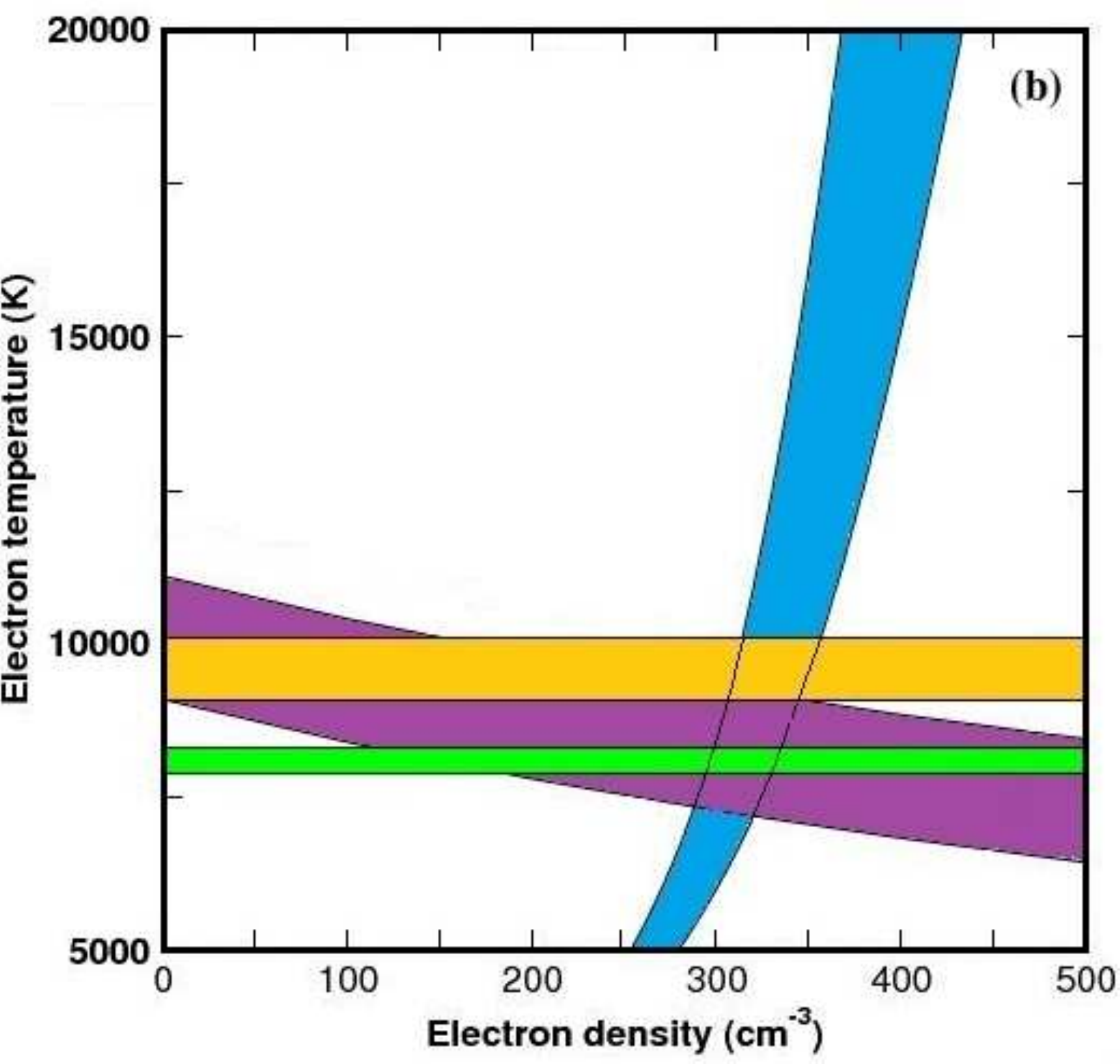}
\includegraphics[width=6cm]{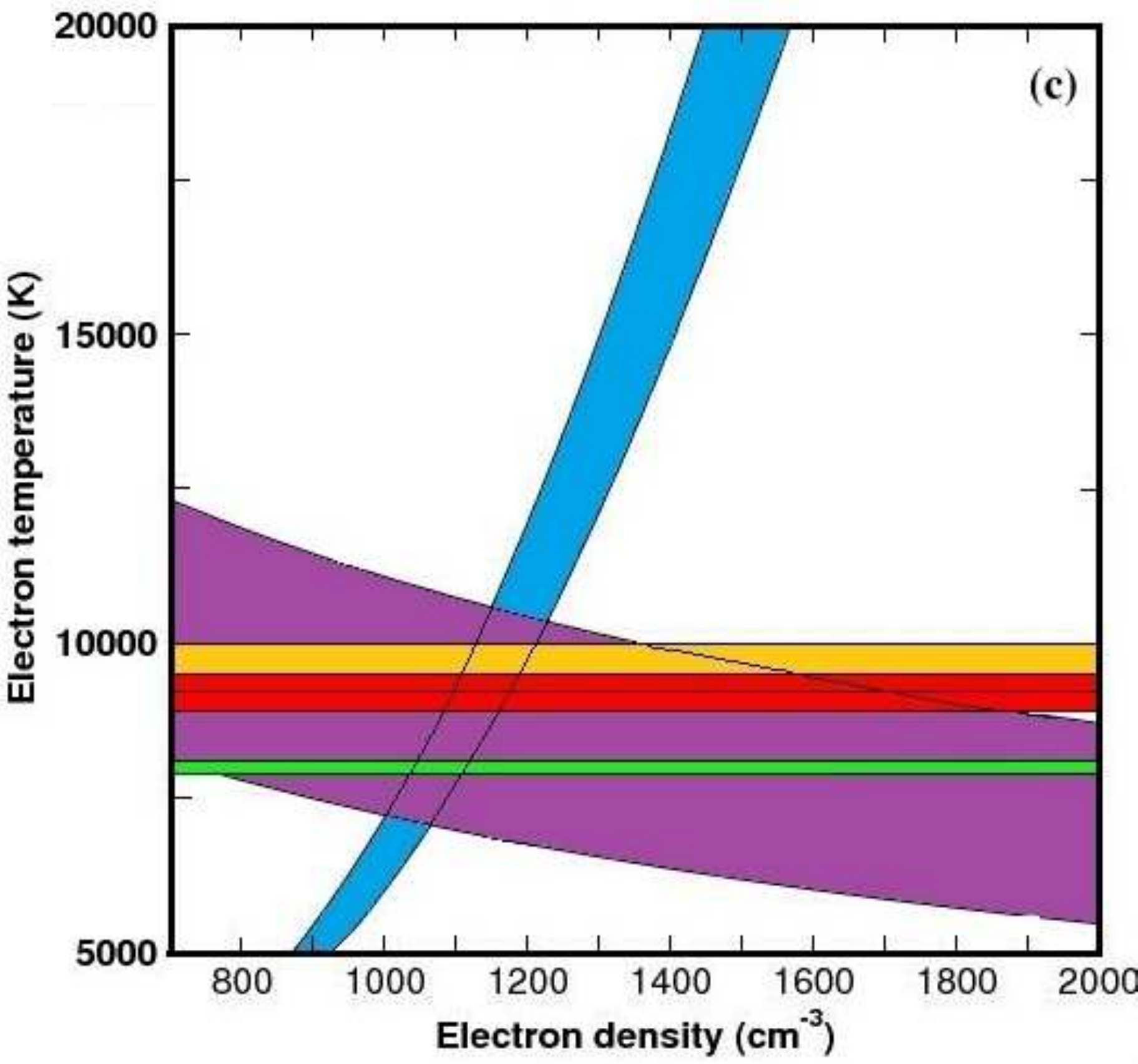}
\includegraphics[width=6cm]{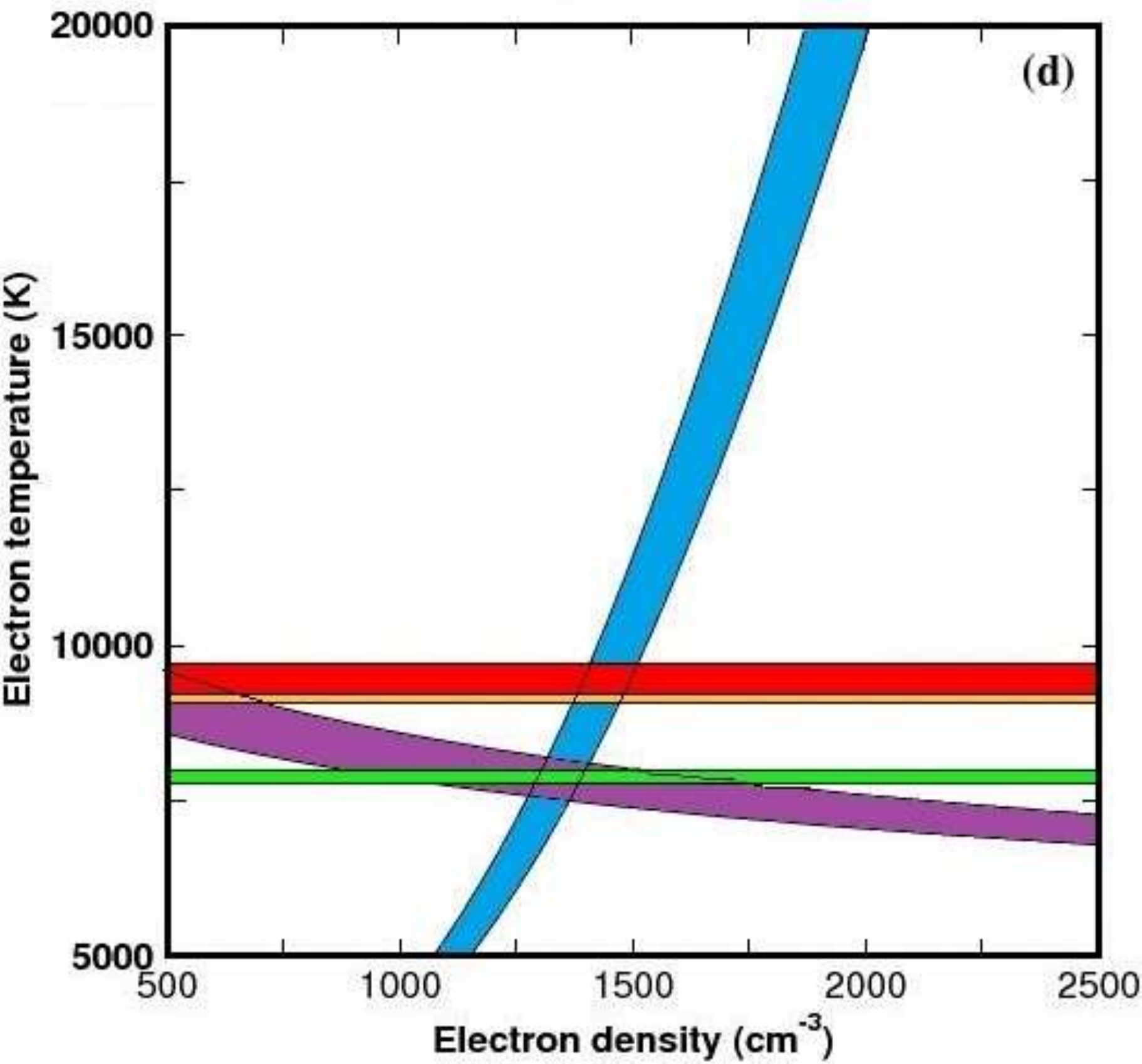}
\includegraphics[width=6cm]{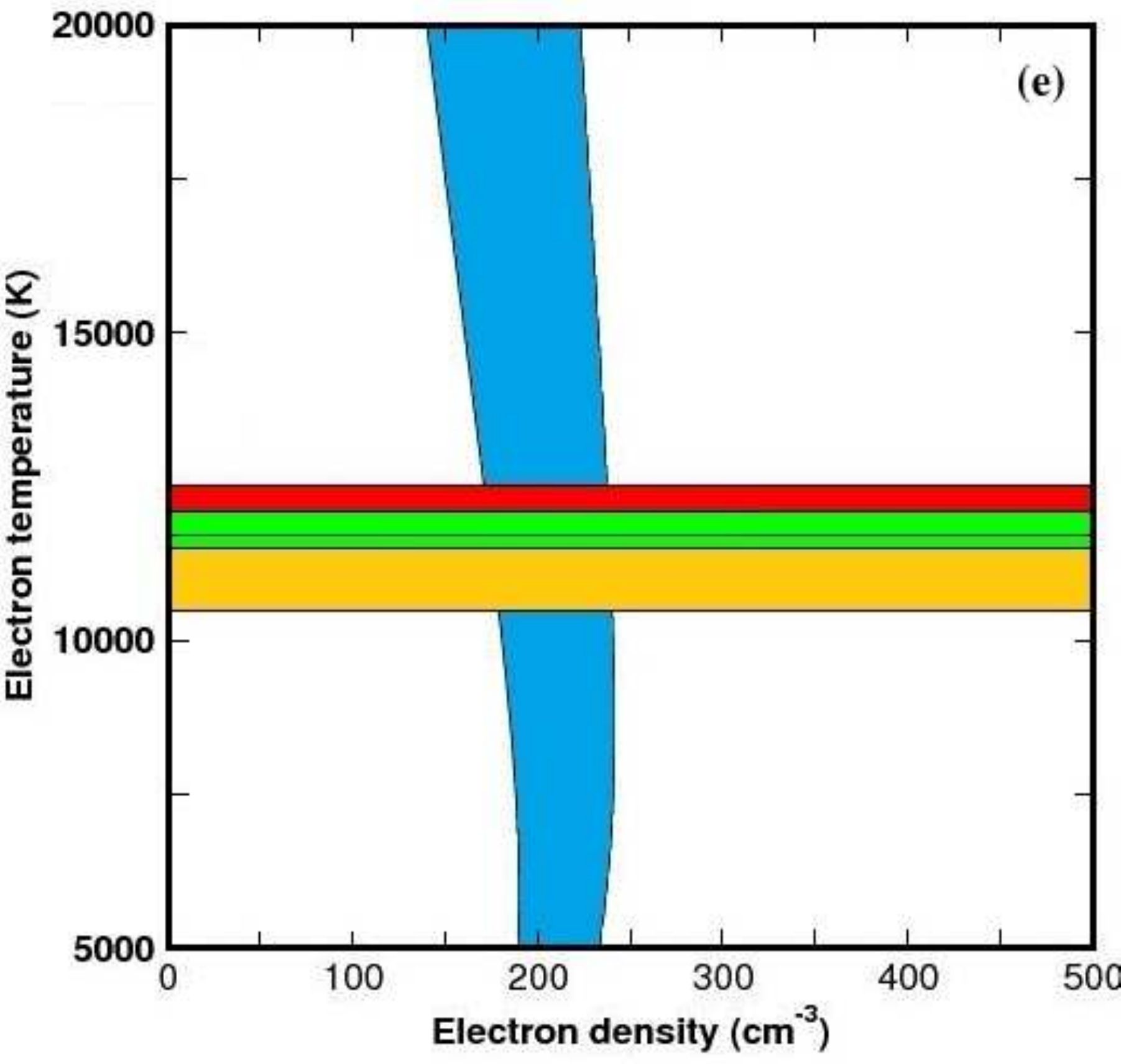}
\caption{T$_e$-n$_e$ diagnostic diagrams of \HII~regions observed with ISIS with direct electron temperatures estimations: (a) S83, (b) S132, (c) S156, (d) S162, 
and (e) S212. Blue bands correspond to n$_e$, purple to t$_e$(\SII), yellow to t$_e$(\NII), red to t$_e$(\OIII), and green to t$_e$(\SIII).}
\label{fig:diagdiag}
\end{figure*} 

Our criterion in the determination of the electron temperature was to derive at least one temperature characteristic for each ionisation degree: t$_e$(\OIII) as the
high-ionisation zone, t$_e($\SIII) as the medium-ionisation zone, and t$_e$(\NII), t$_e$(\SII), or t$_e$(\OII) as the low-ionisation zone. However, not all the necessary 
auroral lines were detected in all the regions (we did not estimate t$_e$ for all the ions). For S206, S298, and S301 no auroral line associated with a low-excitation 
ion was observed and we resorted to the expression
\begin{equation}
t_e(\OII)=\frac{1.2~+~0.002~n_e~+~4.2/n_e}{t_e(\OIII)^{-1}~+~0.08~+~0.003~n_e~+~2.5/n_e},
 \label{eq:teOIIconOIII}
\end{equation}
which relates t$_e$(\OII) with t$_e$(\OIII) \citep{PerezMontero2009}. Equation \ref{eq:teOIIconOIII}  was also used for S127 and S128, in which we inferred t$_e$(\OIII) 
from t$_e$(\OII). In the case of S132, S158, and S255 we estimated t$_e$(\OIII) by using the relation 
\begin{equation}
t_e(\OIII)=\frac{t_e(\NII)}{1.85~-~0.72~t_e(\NII)}
 \label{eq:teOIIIconNII}
\end{equation}
obtained from \citet{PerezMontero2009}. A similar scenario occurs when deriving t$_e$(\SIII) in regions in which \SIII$\lambda$6312 and \SIII$\lambda\lambda$9068,9530 
were not measured simultaneously (S127, S128, S158, S298, S301, and S311). For theses objects we resorted to the expression proposed by \citet{PerezMontero2003} that 
relates  t$_e$(\SIII) with t$_e$(\OIII):
\begin{equation}
t_e(\SIII)=1.19~t_e(\OIII)~-~0.32 .
 \label{eq:teSIIIconOIII}
\end{equation}
Equations \ref{eq:teOIIconOIII} and \ref{eq:teOIIIconNII} are based on photoionisation models, while equation \ref{eq:teSIIIconOIII} was obtained 
with observational data. Electron temperatures estimated with these methods are quoted in Table \ref{table:physical_parameters} labelled with the subscript E. \\

Although the intensities of the necessary auroral lines are relatively reliable, the estimation of the electron temperature is not exempt from problems. 
There are several sources of uncertainty that need to be mentioned because the variations in the temperature would involve offsets in the chemical abundances. 
\citet{Kennicutt2003} compare the consistency of electron temperatures measured from different ions based on observation of \HII~regions in M101. They find that 
the temperatures derived from the \OII$\lambda$7325 line show a large scatter and are nearly uncorrelated with temperatures derived from other ions. The source of 
this disagreement remains unresolved, but they speculate on possible sources such as dielectronic recombination, collisional de-excitation, radiative transfer effects, 
and observational uncertainties. On the other hand, \citet{Binette2012} analyse pairs of \OII~and \SII~temperatures in \HII~regions finding that T$_e$(\OIII) 
appears to be higher than T$_e$(\SIII) in objects with T$_e$(\SIII) lower than 14000~K. They perform photoionisation models to look for an explanation for this 
trend and conclude that models with metallicity inhomogeneities or with shock waves that propagate in the photoionised gas successfully reproduce the observed 
excess in T$_e$(\OIII) temperatures.\\

In order to determine the uncertainties in physical parameters and to analyse possible discrepancies between electron temperatures associated with different ions, we 
performed diagnostic diagrams that show t$_e$ dependences with n$_e$ and vice versa. We represented extreme values of t$_e$ and n$_e$ characterised by the flux ratios 
and their observational errors obtaining diagnostic bands that allow us to visualise the electron temperature structure in the ionised gas. 

Figure \ref{fig:diagdiag} shows the diagnostic diagrams generated for the WHT regions with direct estimation of t$_e$: S83, S132, S156, S162, and S212. 
We show the electron density and its dependence with t$_e$ in blue. Yellow, red, and green bands represent electron temperatures independent of density, respectively t$_e$(\NII), 
t$_e$(\OIII), and t$_e$(\SIII). Finally, we represent the single electron temperature sensitive to density, t$_e$(\SII), in purple. The criterion used to paint the 
parameters is arbitrary and the parameters are represented giving preference to some of them following this order:  
t$_e$(\SIII) - t$_e$(\OIII) - t$_e$(\NII) - t$_e$(\SII) - n$_e$. For this reason, in some regions certain temperatures are not clearly observed because they are hidden 
behind others.

Analysing the density variations, we found that S83, S132, and S212 always show medium densities with values between 150~cm$^{-3}$ and 400~cm$^{-3}$, while S156 and S162 
present higher densities with a wider range of values, from 800~cm$^{-3}$ to 2000~cm$^{-3}$, for a given temperature. In general, the agreement between electron 
temperatures is very good, and it  is possible to appreciate the consistency both for t$_e$ ranges of different ions at estimated n$_e$ and for values presented in 
Table \ref{table:physical_parameters}. The electron temperature associated with \SIII~is the only one that shows lower values than the others in three regions for a fixed 
density; this has been taken into account for the derivation of ionic abundances, as we show in the next section.

\subsection{Direct chemical abundances}\label{sect:abundances}
The ionic chemical abundances of the different species were estimated from the forbidden-to-hydrogen emission line ratios of the strongest available emission lines 
detected in the analysed spectra. We resorted to the functional forms given by \citet{Hagele2008} which are based on the package {\tiny IONIC} of {\tiny IRAF}. 
To determine the single ionised helium abundance we used the equations proposed by \citet{Olive2004}. We checked that results for He$^+$/H$^+$ with this method do not 
differ more than $\sim$3 per cent from values estimated with recent emissivities proposed by \citet{Porter2013}. We also calculated the mean weighted by the errors of the 
five observed \HeI~lines (\HeI$\lambda$4026, \HeI$\lambda$4471, \HeI$\lambda$5875, \HeI$\lambda$6678, and \HeI$\lambda$7065). The ionic abundances of the different elements 
with respect to ionised hydrogen along with their corresponding errors are given in Table \ref{table:abundances}.

To estimate the chemical abundances, electron density and temperature were required. We derived each ionic abundance with its corresponding electron temperature, if it 
was possible. In this way we estimated O$^{+}$ with t$_e$(\OII) , O$^{2+}$ with t$_e$(\OIII), S$^{+}$ with t$_e$(\SII), S$^{2+}$ with t$_e$(\SIII), and N$^{+}$ with t$_e$(\NII). 
In the case of argon we assumed t$_e$(\ArIII)=t$_e$(\SIII) \citep{Garnett1992}, while for iron, helium, and neon the ionic abundances were derived under the assumption 
that t$_e$(\FeIII)=t$_e$(\OIII), t$_e$(\HeI)=t$_e$(\OIII), and t$_e$(\NeIII)=t$_e$(\OIII) \citep{Peimbert1969}. For those regions where not all the electron temperatures 
were derived we assumed an ionisation structure: the adopted criterion was to derive abundances with electron temperatures related with ions with the same excitation degree 
and similar ionisation potential. Nonetheless, we must bear in mind that assuming that temperatures are uniform in a given ionisation zone, our results can be subject to slight 
changes if we consider the presence of temperature inhomogeneities \citep{Esteban2009,MesaDelgado2010,Stasinska2013}. The temperature fluctuation paradigm was proposed 
by \citet{Peimbert1967} and it is characterised by the mean square of the spatial distribution of temperature (the so-called temperature fluctuation parameter t$^2$). Our 
direct chemical abundances are derived using the standard method based on collisional excited lines (CELs), which have a strong dependence on T$_e$; therefore, temperature 
fluctuations can result in an underestimation of the absolute abundances of the ionised gas. This will be considered in the analysis of the chemical gradients.\\

The total abundances were derived by taking into account, when required, the unseen ionisation stages of each element, using the appropriate ionisation correction 
factor (ICF) for each species. The O/H was obtained by directly adding the two ionic abundances (O/H$\sim$O$^{+}$/H$^{+}$+O$^{2+}$/H$^{+}$). The N/O can be 
approximated to N$^{+}$/O$^{+}$ and the N/H ratio as N/H$\sim$(N/O)$\times$(O/H). Due to the spectral range of our observations, no \SIV~line was observed, but -- by taking the 
excitation condition into account -- a relatively important contribution from S$^{3+}$ may be expected in these \HII~regions. Therefore, the total sulfur abundance was inferred 
using the ICF(S$^+$+S$^{2+}$) proposed by \citet{Barker1980}. The total abundances of neon and argon were calculated using the ionisation correction factors 
ICF(Ne$^{2+}$) and ICF(Ar$^{2+}$), respectively, given by \citet{PerezMontero2007}. The total abundance of iron was inferred 
using the ICF(Fe$^{2+}$) proposed by \citet{Rodriguez2004}. Finally, in the case of helium we do not detect the \HeII$\lambda$4686 line (belonging to the observed 
spectral range) in any region and we can assume that He$^{2+}$=0. On the other hand, it is expected that the contribution of the unobservable neutral helium was important. 
In this work, we adopted the ICF(He$^{+}$+He$^{2+}$) proposed by \citet{Peimbert1992} to consider the He$^{0}$ contribution. In Table \ref{table:abundances} we 
present the total direct abundances and the ICFs derived along with their corresponding errors. \\

\subsection{Tailor-made photoionisation models}\label{sect:models}
For those regions without a detection of any auroral temperature-sensitive emission line (e.g. \NII$\lambda$5755, \OIII$\lambda$4363) we were not able to derive chemical 
abundances following the direct method. Instead, it is possible to apply methods based on strong emission line ratios, such as R$_{23}$, O$_3$N$_2$, or S$_{23}$. However, 
most of these methods have been calibrated for massive star-forming complexes whose scales and properties do not match those in our sample of nebulae ionised by single 
stars in the disc of the Galaxy. As an alternative, we can resort to tailor-made photoionisation models covering the observed properties of these objects.  

In Table \ref{table:infoRHII} we showed the spectral type of the dominant exciting stars of the sample. These spectral types were derived by different authors using different 
methods, and great discrepancies among authors in spectral types for the same star are found. For this reason, and in order to work with a self-consistent sample, we decided 
to estimate representative effective temperatures resorting to the $\eta'$ parameter instead of that derived from the spectral type in Table \ref{table:infoRHII}.

The spectral energy distributions (SEDs) of the ionising stars can be obtained from the observed emission line ratios using appropriate diagnostic diagrams that yield a 
stellar effective temperature. This is the case of the $\eta$ parameter \citep{Vilchez1988}, defined as
\begin{equation}
\mathrm{ \eta=\frac {O^{+}/O^{2+}} {S^{+}/S^{2+}} },
 \label{eq:eta}
\end{equation}
which is sensitive to the hardness of the ionising spectral energy distribution. When the ionic abundances cannot be derived, it is possible to use the $\eta'$ parameter 
based on emission lines:
\begin{equation}
\mathrm{ \eta'=\frac {\OII 3727/ \OIII 4959,5007} {\SII 6717,6731/\SIII 9069,9532} }.
 \label{eq:etap}
\end{equation}

\begin{figure}  
\centering
\includegraphics[width=7cm]{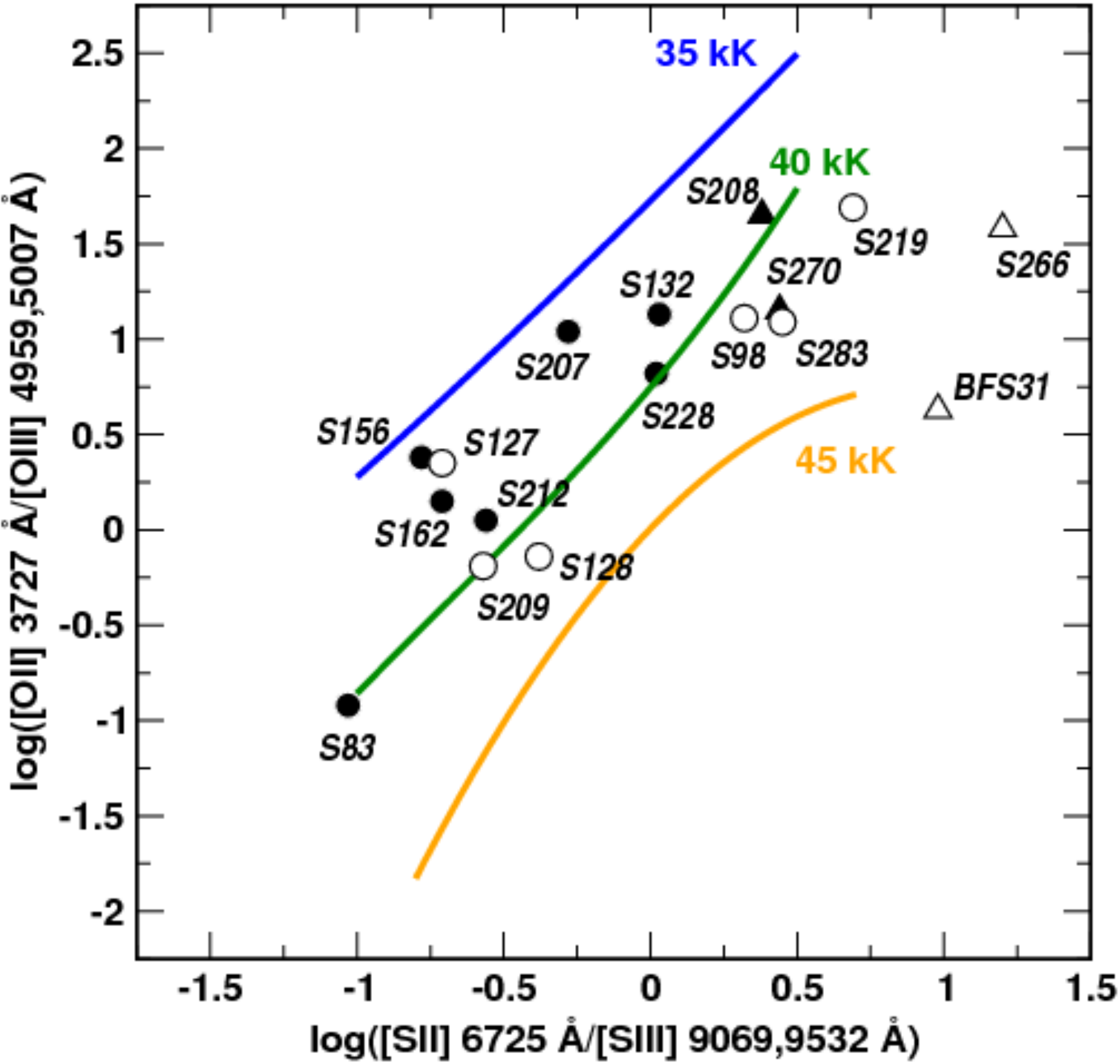}
\caption{\SII/\SIII~versus \OII/\OIII~in logarithm units compared with a grid of WM-Basic models at Z=0.2 Z$_\odot$. The solid lines 
represent models with the same effective temperature as indicated. Circles show \HII~regions with reliable line emission measurements and triangles regions with upper 
limit intensities. Filled symbols represent data from our own sample (observed regions) while open symbols represent regions taken from the literature.}
\label{fig:teff}
\end{figure}

Although this parameter mainly correlates with the stellar effective temperature, it also has an additional dependence on electron temperature and, hence, on chemical 
abundance. We then used the calibration of the $\eta'$ parameter presented by \citet{PerezMontero2014} based on fittings to  models with WM-Basic \citep{Pauldrach2001} 
single stars at a metallicity Z=0.2 Z$_\odot$ close to the expected values for this subsample. The $\eta'$ criterion limits the use of models when deriving chemical abundances 
to those \HII~regions with the four involved emission lines. All the \HII~regions with measures of \OII$\lambda$3727,  \OIII$\lambda$5007, \SII$\lambda\lambda$6717,6731, 
and \SIII$\lambda$9069 are shown along with the models used to derive the effective temperature in Fig. \ref{fig:teff}.\\

\setcounter{table}{5}
\begin{table}
\caption{Obtained properties from the resulting tailor-made photoionisation models.}
\label{table:models} 
\centering 
\begin{tabular}{l c c c c c }
\\
\hline
		&	\Teff 	&	$\log$ L(\Hb) 	&	Radius &	12+log(O/H)	&	log(N/O)	\\
		&	(kK)	&	 (erg/s)	&	(pc)		&			&			\\
\hline\hline
S83$^*$	&	40	&	34.15		&	0.1		&	8.16	&	-0.95	\\
S98			&	41	&	35.02		&	4.4		&	8.16	&	-1.06	\\
S127$^*$	&	37	&	34.33		&	0.3		&	8.29	&	-1.34	\\
S128$^*$	&	41	&	34.57		&	0.7		&	8.21	&	-1.00	\\
S132$^*$	&	38	&	33.02		&	0.1		&	8.32	&	-0.98	\\
S156$^*$	&	36	&	34.11		&	0.1		&	8.26	&	-0.95	\\
S162$^*$	&	37	&	33.57		&	3.0		&	8.30		&	-0.86	\\
S207		&	40	&	34.23		&	2.0		&	8.13	&	-1.06	\\
S208		&	39	&	33.90		&	1.7		&	$>$8.31	&	$<$-0.96	\\
S212$^*$	&	39	&	34.76		&	4.1		&	8.15	&	-1.06	\\
S219		&	42	&	33.35		&	3.7		&	$>$7.92	&	$<$-0.86	\\
S228		&	40	&	34.33		&	1.4		&	8.23	&	-0.92	\\
S266		&	42	&	33.21		&	0.6		&	$>$8.11	&	$<$-0.78	\\
S270		&	42	&	34.53		&	1.1		&	$>$7.72	&	$<$-0.66	\\
S283		&	42	&	33.17		&	1.0		&	8.20	&	-1.19	\\
\hline
\end{tabular}
\begin{list}{}{} \footnotesize{
\item {$^*$} \HII~regions with chemical abundances also estimated with the direct method.
}
\end{list}
\end{table}

\begin{figure}  
\includegraphics[width=\columnwidth]{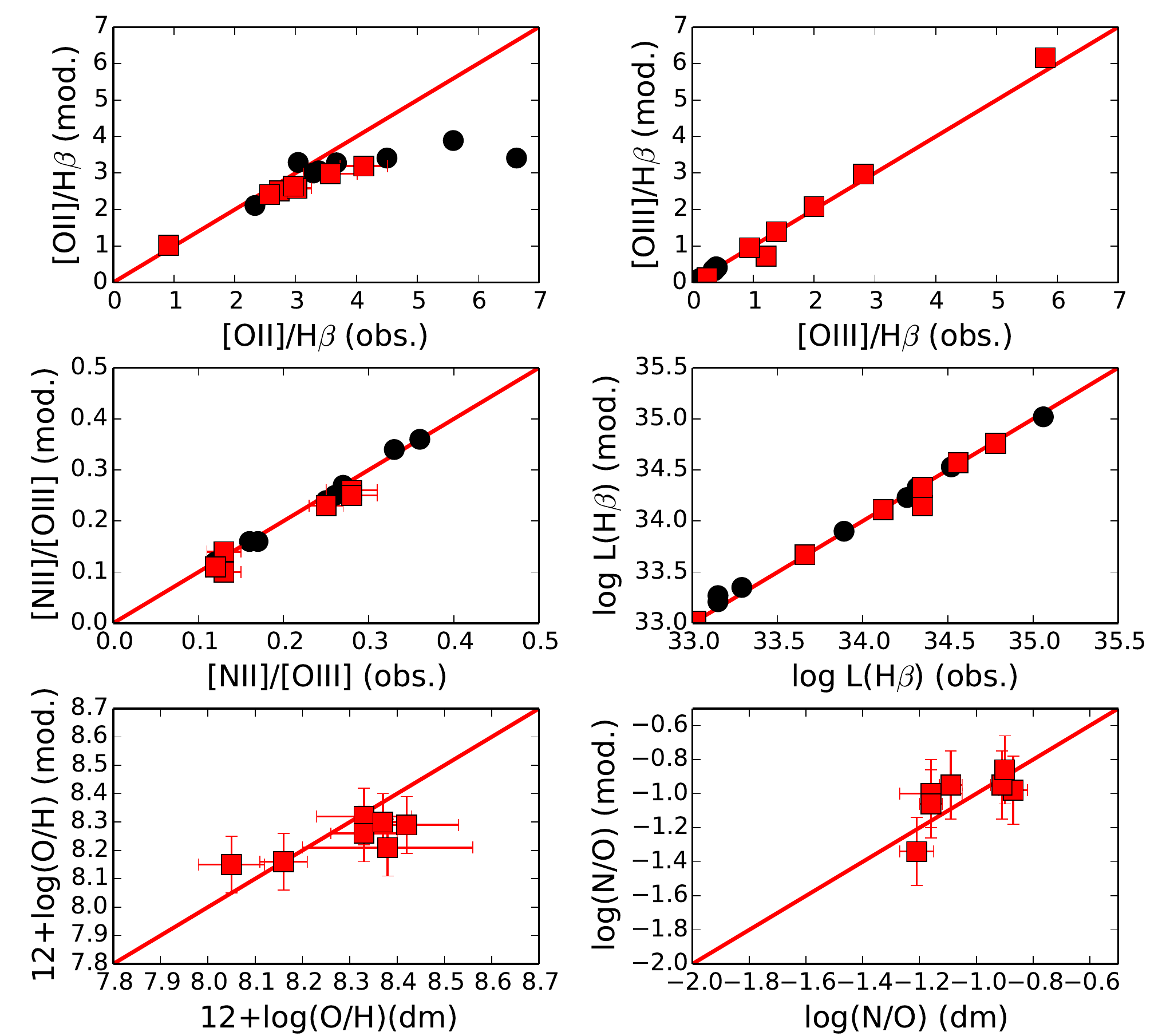}
\caption{Upper and middle rows show a comparison between observed quantities and the found solutions of the tailor-made models, while in the last row we compare the chemical 
abundances obtained from models with those derived with the direct method. Red squares represent \HII~regions with direct estimations of T$_e$ and black circles represent 
regions without a detection of any auroral temperature-sensitive emission line.}
\label{fig:models}
\end{figure}

Once the ionising SEDs were established, we used the photoionisation code Cloudy v13.03  \citep{Ferland2013} using as input conditions a constant density in the gas 
according to the value derived from the ratio of \SII~emission lines and a dust-to-gas ratio with a standard value of 2$\times$10$^{-3}$. All models have a radiation-bounded 
geometry and the stopping criterion is that 99\% of the H atoms are ionised. The models for each region are repeated in an iterative process varying Q(H), the inner radius, 
the number of hydrogen ionising photons, the abundance of all elements scaled to oxygen, and the  nitrogen-to-oxygen ratio that most closely reproduces the 
observed \OII$\lambda$3727/\Hb, \OIII$\lambda$5007/\Hb, \NII$\lambda$6584/\OII$\lambda$3727, and the H$\beta$ luminosity. In the S208, S219, S266, and S270 regions, 
the \OIII$\lambda$5007 line is not detected or it has a very weak flux, so we considered an upper limit of this line two times the uncertainty associated with the continuum 
at the same wavelength. The typical number of iterations before a solution is found is around 50, with the exception of S203 and S209 where no reliable solutions were found. 
In the former region, the relatively very high \OIII/\Hb~ratios is possibly due to a density-bounded geometry and in the latter to possible aperture effects.

The obtained properties from the resulting models, including L(\Hb), external radius, oxygen abundance, and N/O ratios are summarised in Table \ref{table:models}. 
Figure \ref{fig:models} shows the agreement between the four observed quantities and the found solutions. With the exception of \OII/\Hb~in S98 and S283, errors in the accuracy 
of the adjustments are less than 1 per cent for H$\beta$ luminosity, 5 per cent for \OIII/\Hb~and \NII/\OII, and 10 per cent for \OII/\Hb. In S98 and S283 the observed very high 
value of \OII/\Hb~is possibly due to some contamination from the diffuse gas. In the last row of Fig. \ref{fig:models} we also show the comparison between chemical 
abundances obtained from models (Table \ref{table:models}) with those derived with the direct method (Table \ref{table:abundances}). As can be seen, the results from the two 
methods are in good agreement showing that the two treatments are consistent.

\subsection{Relation between radio and optical electron temperatures}\label{sect:tradio}
In this subsection we perform a comparative analysis of the optical and radio electron temperatures to derive a relation between them. Since there is a large sample of nebulae 
widely distributed across the Galactic disc with estimations of electron temperatures based on radio recombination lines (RRL) and continuum data, this relation could be 
used to obtain the optical temperature and to calculate chemical abundances in those cases without detectable optical auroral lines.

Most of the observed emission lines in ionised nebulae are collisionally excited and their intensities depend exponentially on temperature. This temperature can be determined 
from appropriate line ratios. However, these ratios involve the detection and measurement of auroral lines which are intrinsically weak and remain undetected in many objects.  
On the other hand, most of the radio emission observed from \HII~regions is continuum radiation produced by free-free thermal bremsstrahlung in the plasma. At high frequencies 
the nebular gas is optically thin, and the ratio between the brightness temperature of a RRL and that of a free-free emission continuum depends on the radio frequency and 
the gas temperature. Thus, the observed RRL-to-continuum ratio can be used to estimate the electron temperature of the \HII~regions \citep{Rohlfs2000}. Because the RRLs are 
not obscured by interstellar dust, relatively faint \HII~regions at large distances from the Sun can be detected, allowing radio electron temperatures t${_e^*}$ to be 
estimated more easily than with optical auroral lines. \\

\begin{figure}  
\centering
\includegraphics[width=\columnwidth]{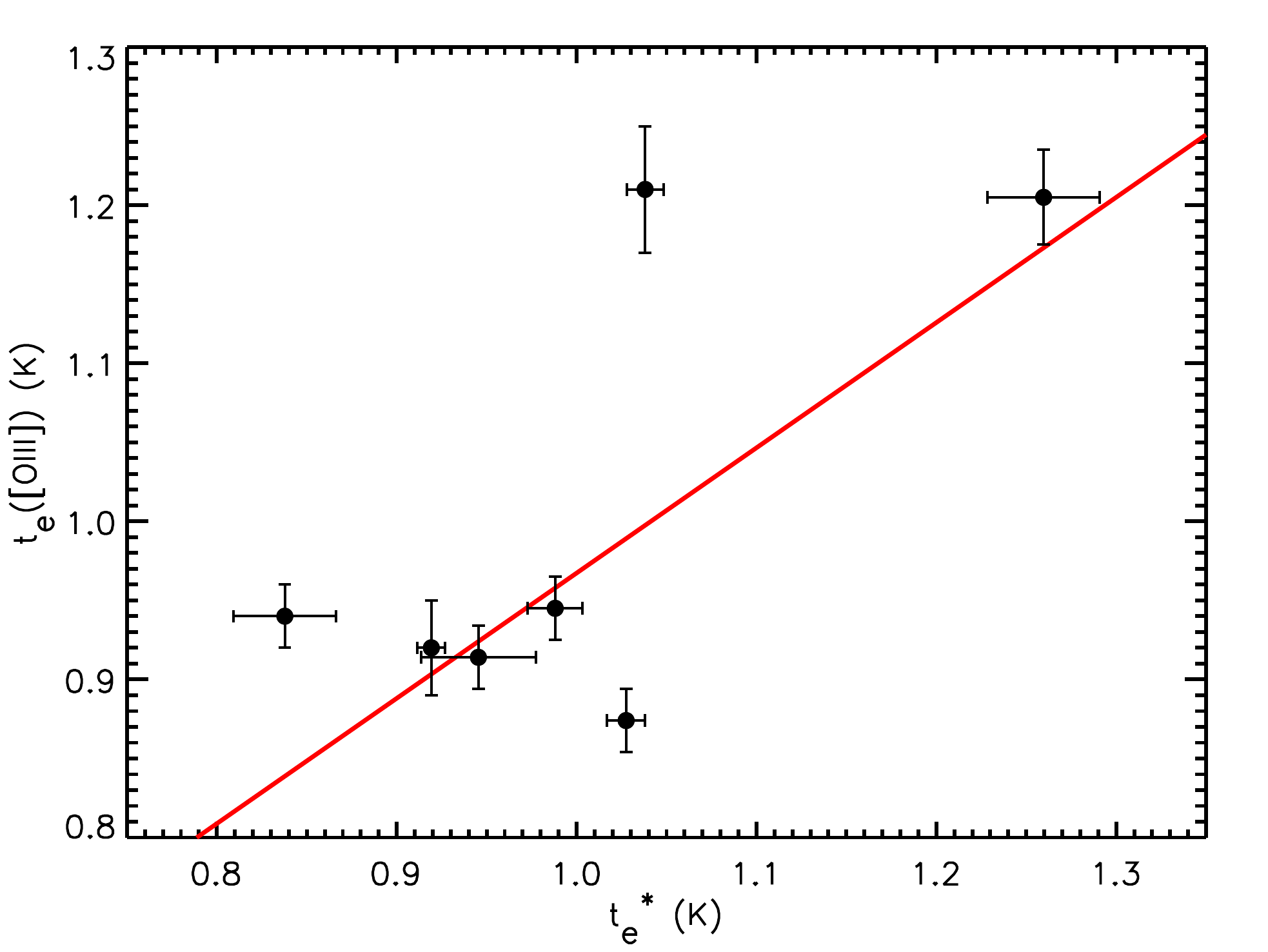}
\caption{Comparison of optical electron temperatures derived from the \OIII~lines, t${_e}$(\OIII), with those derived from the radio recombination lines, t${_e^*}$. The red 
solid line represents the fit performed weighted by errors.}
\label{fig:TeTr}
\end{figure}

With the intention of establishing a relation to convert  t${_e^*}$  into optical electron temperatures t${_e}$ which are relevant in abundance determinations, we 
first checked the literature for studies with reliable estimations of radio temperatures (see Col. 9 of Table \ref{table:infoRHII}). Later, we compared these 
values with our optical electron temperatures derived from the \OIII~lines (see Col. 4 of Table \ref{table:physical_parameters}) for \HII~regions in common. 
Figure \ref{fig:TeTr} shows optical electron temperatures, t${_e}$(\OIII), derived in this work as a function of the radio temperatures, t${_e^*}$, adopted from the literature. 
Each point represents an \HII~region with radio temperatures being a weighted mean of determinations from  \citet{Quireza2006} and \citet{Balser2011}. Despite the reduced 
number of regions, the obtained relation is described by the weighted fit
\begin{equation}
\mathrm{t{_e}}(\mathrm{\OIII}) =(0.175\pm 0.080)+(0.792\pm0.081)\times \mathrm{t_e^*}
\label{eq:TrTe}
\end{equation}

\begin{table}[t!]
\caption{Electron temperatures, electron densities, and ionic and total chemical abundances obtained from the t${_e}$-t$_e^*$ relation.} 
\label{table:Tr_abundances} 
\centering 
\begin{tabular}{l  c c c }
\\
\hline
	 &		S98	 &				S209			 &		S228		\\	
\hline\hline
t$_e^*$	(10$^4$ K) &	1.08	$\pm$ 	0.02	&	1.06	$\pm$ 	0.01	&	0.94	$\pm$ 	0.01	\\
t$_e$(\OIII) (10$^4$ K)	&	1.03	$\pm$ 	0.12	&	1.02	$\pm$ 	0.12	&	0.92	$\pm$ 	0.11	\\
t$_e$(\NII) (10$^4$ K)	&	1.10	$\pm$ 	0.07	&	1.09	$\pm$ 	0.07	&	1.02	$\pm$ 	0.07	\\
t$_e$(\SIII) (10$^4$ K)	&	0.91	$\pm$ 	0.14	&	0.89	$\pm$ 	0.14	&	0.78	$\pm$ 	0.13	\\
n$_e$ (cm$^{-3}$) 	&	$<$100 			&	423	$\pm$ 	49	&	233	$\pm$ 	12	\\
													\\
12+log(O$^+$/H$^+$)	&	8.23	$\pm$ 	0.12	&	7.84	$\pm$ 	0.12	&	8.13	$\pm$ 	0.13	\\
12+log(O$^{2+}$/H$^+$) 	&	7.06	$\pm$ 	0.21	&	7.94	$\pm$ 	0.17	&	7.27	$\pm$ 	0.19	\\
12+log(S$^+$/H$^+$) 	&	5.94	$\pm$ 	0.08	&	5.47	$\pm$ 	0.08	&	6.07	$\pm$ 	0.07	\\
12+log(S$^{2+}$/H$^+$)	&	6.14	$\pm$ 	0.18	&	6.43	$\pm$ 	0.17	&	6.51	$\pm$ 	0.18	\\
12+log(N$^+$/H$^+$) 	&	7.21	$\pm$ 	0.08	&	6.70	$\pm$ 	0.08	&	7.26	$\pm$ 	0.08	\\
log(N$^+$/O$^+$) 	&	-1.01	$\pm$ 	0.07	&	-1.10	$\pm$ 	0.08	&	-0.85	$\pm$ 	0.06	\\
(He$^+$/H$^+$) 5875 	&	-	 		&	0.08	$\pm$ 	0.01	&	-	 		\\
(He$^+$/H$^+$) 6678	&	-	 		&	0.08	$\pm$ 	0.03	&	0.09	$\pm$ 	0.01	\\
(He$^+$/H$^+$) 7065	&	-	 		&	0.07	$\pm$ 	0.02	&	0.07	$\pm$ 	0.01	\\
(He$^+$/H$^+$) 	&	-	 		&	0.08	$\pm$ 	0.01	&	0.08	$\pm$ 	0.01	\\
													\\
ICF(S$^+$,S$^{2+}$)	&	1.00	$\pm$ 	0.01	&	1.11	$\pm$ 	0.07	&	1.00	$\pm$ 	0.01	\\
ICF(He$^+$) 	&	-			&	1.10	$\pm$ 	0.04	&	1.36	$\pm$ 	0.16	\\
													\\
12+log(O/H) 	&	8.26	$\pm$ 	0.12	&	8.19	$\pm$ 	0.11	&	8.18	$\pm$ 	0.12	\\
12+log(S/H) 	&	6.36	$\pm$ 	0.12	&	6.52	$\pm$ 	0.16	&	6.65	$\pm$ 	0.13	\\
12+log(N/H) 	&	7.24	$\pm$ 	0.19	&	7.05	$\pm$ 	0.19	&	7.29	$\pm$ 	0.19	\\
log(N/O) 	&	-1.01	$\pm$ 	0.07	&	-1.10	$\pm$ 	0.08	&	-0.85	$\pm$ 	0.06	\\
He/H 	&	-			&	0.09	$\pm$ 	0.01	&	0.11	$\pm$ 	0.02	\\

\hline
\end{tabular}
\end{table}

This result differs from that established by \citet{Shaver1983}, who found that the weighted average of the ratio of these temperatures is t${_e}$(\OIII)/t${_e^*}$=1.06$\pm$ 0.02, with the result that the slope derived here is smaller than their value. Their relation was obtained from only six \HII~regions located close to the Sun, but we have not found other studies in the literature that can be compared with our results. \\

We applied the relation obtained  to those regions of the sample without direct estimations of optical temperatures, but with information of radio temperature from the 
literature (S98, S209, and S228). Once t${_e}$(\OIII) was estimated for each region using Eq. \ref{eq:TrTe}, we calculated temperatures associated with other ions with 
different ionisation degrees (t${_e}$(\NII) with Eq. \ref{eq:teOIIIconNII} and t${_e}$(\SIII) with Eq. \ref{eq:teSIIIconOIII}). Finally, we inferred chemical abundances 
following the same procedure as for regions with observable auroral lines (Sect. \ref{sect:abundances}). Electron temperatures, ionic abundances, and chemical abundances 
derived with this method are shown in Table \ref{table:Tr_abundances}. For two regions, S98 and S288, chemical abundances were also estimated resorting to the photoionisation 
models described in the previous section; as can be seen (Tables \ref{table:models} and \ref{table:Tr_abundances}), both derivations are consistent within errors.

%SECTION 4 
%%%%%%%%%%%%%%%%%%%%%%%%%%%%%%%%%%%%%%%%%%%%%%%%%%%%%%%%%%%%%%%%%%%%%%%%%%%%%%
\section{Discussion} \label{sect:discussion}

\subsection{Galactocentric distances of the \HII~regions } \label{sect:distances}

\begin{figure}  
\centering
\includegraphics[width=6cm]{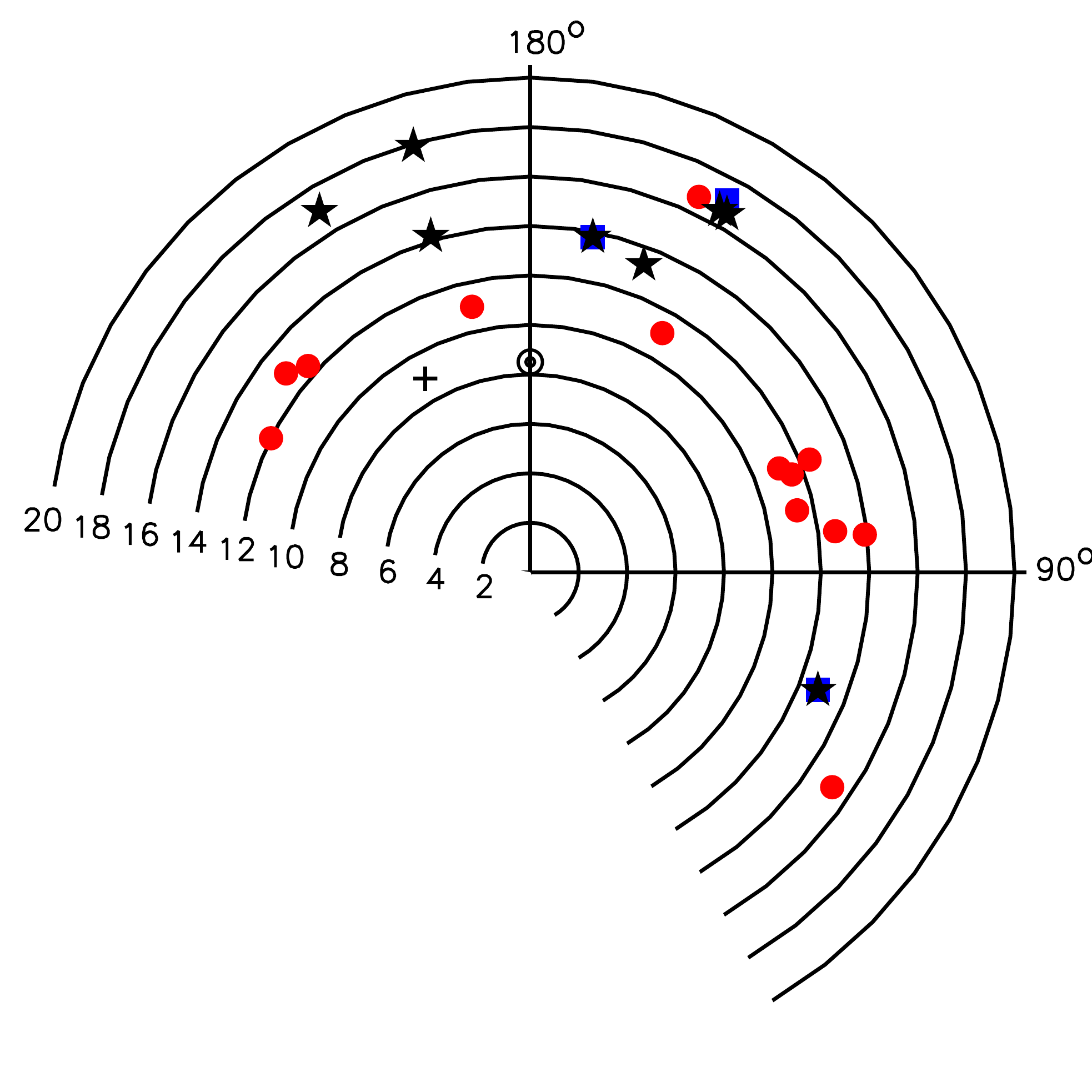}
\caption{Distribution of the \HII~regions analysed in this study projected onto the Galactic plane. The numbers close to the circles show distances to the Galactic centre in 
kpc. Red circles denote regions with abundances estimated with direct methods, black squares represent results from tailor-made models, and blue squares plot the abundances 
obtained from the t$_e$- t$_e^*$ relation. Regions with chemical abundances estimated with two methods are represented twice. The Sun and M\,42 are also represented 
at R$_G$=8.5~kpc (black solar symbol) and R$_G$=8.95~kpc (black cross), respectively.}
\label{fig:positions}
\end{figure} 

The study of abundance gradients in galaxies is strongly affected by the uncertainty in galactocentric distance: it can affect the slope and, therefore, the interpretation 
of the results. The main difficulty in establishing the parameters shown in Cols. 4 and 5 of Table \ref{table:infoRHII} was to decide what Galactocentric radius of 
the \HII~regions to adopt in this work. We searched in the literature for the most accurate distances for our sample, and found several previous works which used different 
methods: stellar photometry, kinematical, or mixed techniques. In order to generate a self-consistent sample and to minimise the effects of inaccuracies in distances, we 
selected Galactic and heliocentric distances from the same catalogue for all the regions whenever possible.

Taking these considerations into account, in a first attempt, we adopted the Galactocentric distances from \citet{Caplan2000} and \citet{Quireza2006}. From these  
works we obtained information about R$_G$ and \dsol~for 17 \HII~regions (77$\%$ of the sample). Both groups of authors use the same method, deriving distances kinematically 
from the observed nebular recombination line local standard of rest (LSR) velocity, assuming the \citet{Brand1993} Galactic rotation curve and placing the Sun at a 
Galactocentric distance of \Rsol=8.5~kpc orbiting the Galactic centre at a LSR circular velocity of $\theta$=220~km~s$^{-1}$. The two works complement each other well since 
they give distances for different regions or similar estimations for those regions listed in both works. Only in the case of S212 do \citet{Caplan2000} and 
\citet{Quireza2006} differ in the derived distances, estimating R$_G$=14.8~kpc and 16.7~kpc, respectively. We adopted the latter because it agrees with the value 
proposed by \citet{Balser2011}, whose estimation will be used for other regions (see below). 

Five regions of our sample do not have distances estimated by \citet{Caplan2000} or \citet{Quireza2006}. For these regions, we decided to use works that 
derive distances with  the same methodology. To this end, we adopt distances derived by \citet{Fich1991} for S98, S266, S270, and S283 and by \citet{Balser2011} for S98. 
These five regions are not individually catalogued in any other study of kinematical distances. Both \citet{Fich1991} and \citet{Balser2011} derive kinematic distances 
with \Rsol=8.5~kpc and $\theta$=220~km~s$^{-1}$, although \citet{Fich1991} use a flat rotation curve instead a curve from \citet{Brand1993}.

Here we have thus chosen to use only these kinematic determinations of the R$_G$ of our sources, obtaining a sample with consistent Galactocentric radius. 
Figure \ref{fig:positions} shows the distribution of the 22 \HII~regions with derived chemical abundances projected onto the Galactic plane and plotted as function of the 
Galactic longitude and Galactic radius. They cover the anticentre range of distances from $\sim$11~kpc to $\sim$18~kpc.

\subsection{Abundance gradients of O/H, N/H, N/O, S/H, Ar/H, and He/H towards the Galactic anticentre}\label{sect:gradients}

\begin{table}[t!]
\caption{Final derived abundance gradients towards the Galactic anticentre. Linear least-squares fits weighted by abundances errors computed using all the data (three methods) 
and using only direct abundances are presented separately. Abundances are in units of dex and Galactocentric radius in units of kpc. } 
\label{table:gradientes} 
\centering 
\renewcommand{\arraystretch}{1.5}
\begin{tabular} {l  c } 
\\
\hline
Data used	 & Final abundance gradients		\\	
\hline\hline
\multicolumn{2}{c}{12+log(O/H)} \\ 
\hline
All data & (9.006 $\pm$ 0.112) - (0.053 $\pm$ 0.009) R$_G$\\ 
Direct & (9.113 $\pm$ 0.139) - (0.061$\pm$ 0.011 ) R$_G$\\
\hline
\multicolumn{2}{c}{12+log(N/H)} \\ 
\hline
All data &(8.260 $\pm$ 0.258) - (0.080 $\pm$ 0.019) R$_G$\\
Direct & (8.331 $\pm$ 0.274 ) - (0.085 $\pm$ 0.021) R$_G$\\
\hline
\multicolumn{2}{c}{log(N/O)} \\
\hline
All data & - (0.478 $\pm$ 0.086) - (0.041 $\pm$0.006) R$_G$\\
Direct & - (0.399 $\pm$ 0.095) - (0.047 $\pm$ 0.007) R$_G$\\
\hline 
\multicolumn{2}{c}{12+log(S/H)} \\
\hline
All data &(8.162 $\pm$ 0.088) - (0.106 $\pm$ 0.006) R$_G$\\
 Direct &(8.194 $\pm$ 0.089) - (0.108 $\pm$ 0.006) R$_G$\\
\hline
\multicolumn{2}{c}{12+log(Ar/H)} \\
\hline
All data &(7.178  $\pm$ 0.073) - (0.074 $\pm$ 0.006) R$_G$\\
\hline
\multicolumn{2}{c}{He/H} \\
\hline
All data &(0.0968 $\pm$ 0.0259) - (0.0005 $\pm$ 0.0019) R$_G$\\
Direct &(0.0982 $\pm$ 0.0299) - (0.0007 $\pm$ 0.0022) R$_G$\\
\hline
\end{tabular}
\renewcommand{\arraystretch}{1}
\end{table}

\begin{table*}[t!]
\caption{Chemical abundance gradients derived by other authors in studies with Galactic \HII~regions. We show the slope, d(log(X/H))/dR$_G$, for each element X in units 
of dex~kpc$^{-1}$.} 
\label{table:grad_literature}
\centering 
\resizebox{\textwidth}{!}{ 
\begin{tabular}{l c c c c c c}
\\
\hline
\\
Reference & Spectral range & R$_G$ range (kpc) & O/H & N/H & S/H &N/O\\
\hline \hline \\
\citet{Shaver1983} & radio+optical & 3.5-13.7 & -0.07 & -0.09 & -0.01 & - \\
\citet{Simpson1995} & radio+infrared & 0.1-10.2 & - & -0.10 & -0.07 & -0.04 \\
\citet{Vilchez1996} $^{[a]}$ & optical & 12.4-18.0 & -0.028/-0.036/-0.051 & 0.043/-0.009/0.002 & -0.01/-0.041/-0.013 & - \\
\citet{Afflerbach1997} & radio+infrared & 0-11.4 & -0.064 & -0.072 & -0.063 & - \\
\citet{Rudolph1997} & radio+infrared & 12.9-17.0 & - & -0.111 & -0.079 & - \\
\citet{Deharveng2000} $^{[b]}$ & optical & 6.5-17.7 & -0.0473/-0.0395 & - & - & - \\
\citet{Esteban2005} $^{[c]}$ & optical (RLs) & 6.3-10.4 & -0.044 & - & - & - \\
\citet{Quireza2006} & radio & 0.1-16.9 & -0.043 & - & - & - \\
\citet{Rudolph2006} $^{[d]}$ & radio+infrared & 0-18 & -0.060/-0.041 & -0.071/-0.085 & -0.046/-0.042 & +0.004/-0.034 \\
\citet{Balser2011} $^{[e]}$ & radio & 4.7-21.9 & -0.0383/-0.0446 & - & - & - \\
This work$^{[f]}$   &   optical  &  11-18   &  -0.053/-0.061  &  -0.080/-0.085  &  -0.106/-0.108  & -0.041/-0.047 \\
\\
\hline
\end{tabular}
} 
\begin{list}{}{} \footnotesize{
\item {$^{a}$} Abundances derived from direct t$_e$. / Abundances derived from model-dependent methods. / All the estimations.
\item {$^{b}$} Completed sample. / Regions with direct t$_e$.
\item {$^{c}$} Chemical abundances derived from recombination lines (RLs) with t$^2\neq$0.
\item {$^{d}$} Optical data. / Far infrared data.
\item {$^{e}$} GBT sample. / Green Bank sample.
\item {$^{f}$} All data (three methods). / Only direct abundances.}
\end{list}
\end{table*}

\begin{figure}  
\centering
\includegraphics[width=\columnwidth]{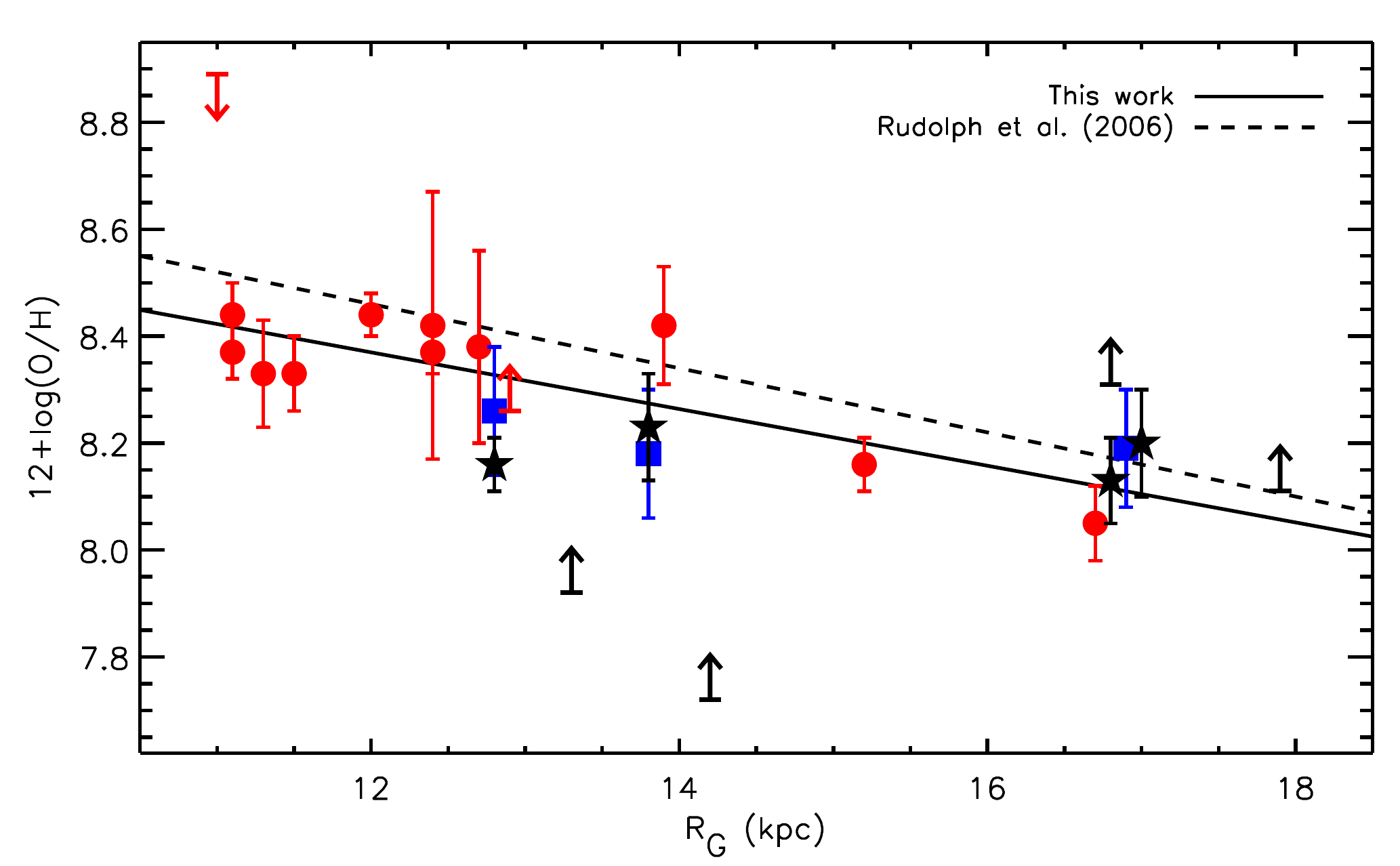}
\caption{Radial distribution of 12+log(O/H) (in dex) plotted versus Galactocentric radius (R$_G$ in kpc). Red circles denote regions with abundances estimated with 
direct methods, black squares represent results from tailor-made models, and blue squares plot the abundances obtained from t$_e$- t$_e^*$ relation. 
Regions with abundances that are upper or lower limits are represented by arrows. The solid line represents the least-squares weighted fit performed to all the data (except 
limits abundances), while the extrapolated gradient from \citet{Rudolph2006} is represented by a dashed line.}
\label{fig:OHgrad}
\end{figure}

\begin{figure}  
\centering
\includegraphics[width=\columnwidth]{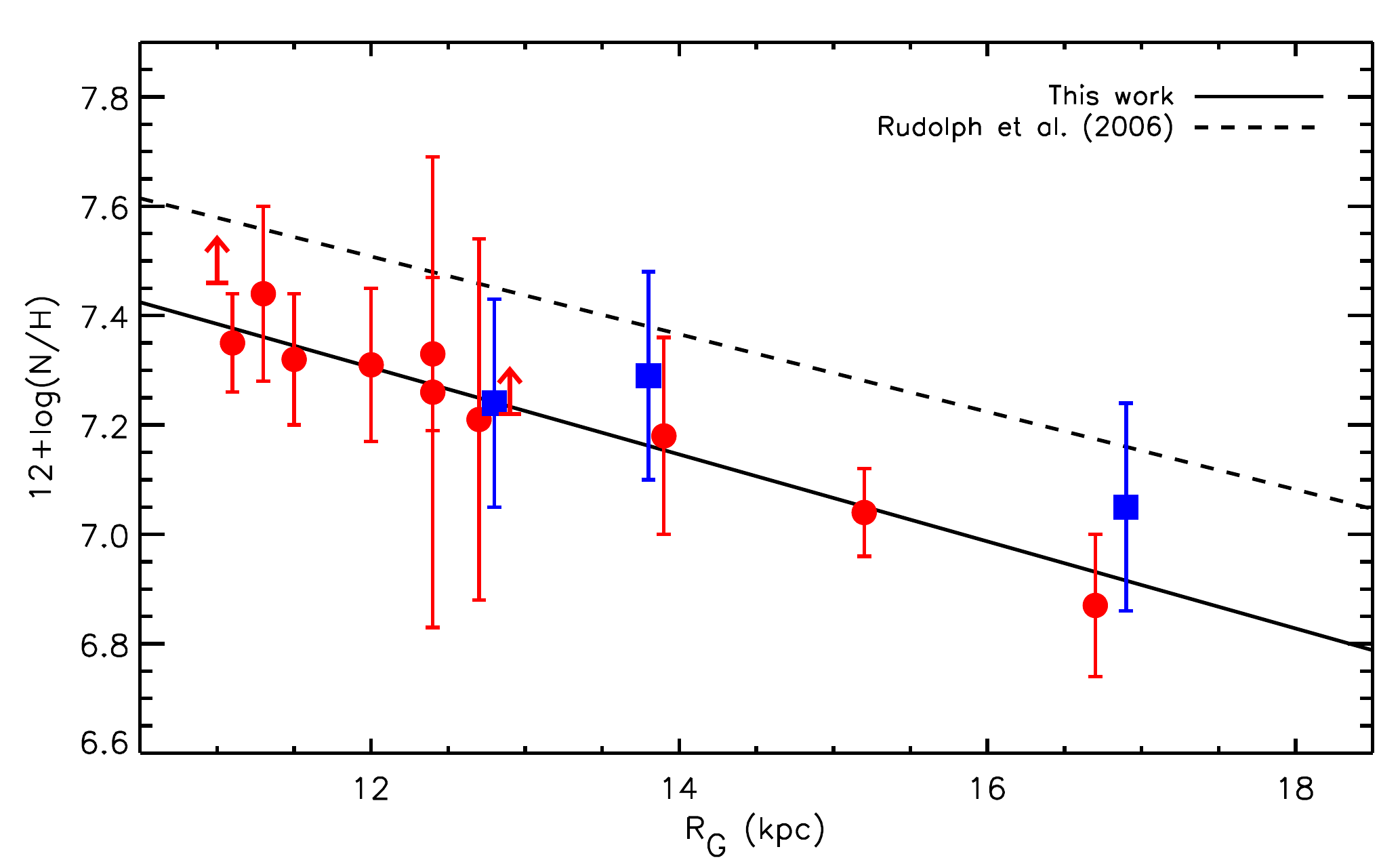}
\caption{Radial distribution of 12+log(N/H) plotted versus Galactocentric radius. Symbols and colours are as in Fig. \ref{fig:OHgrad}.}
\label{fig:NHgrad}
\end{figure}

\begin{figure}  
\centering
\includegraphics[width=\columnwidth]{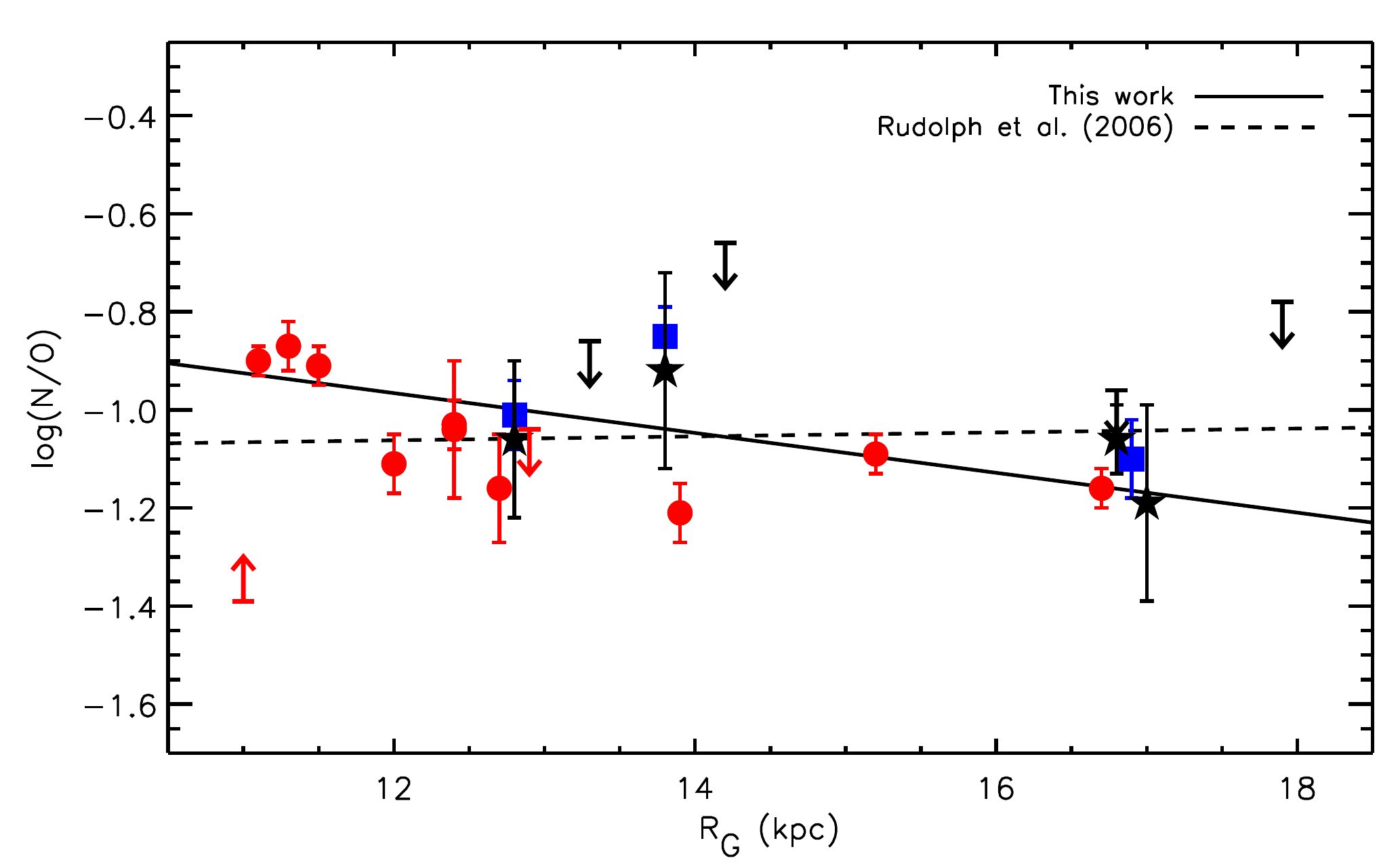}
\caption{Radial distribution of log(N/O) plotted versus Galactocentric radius. Symbols and colours are as in Fig. \ref{fig:OHgrad}.}
\label{fig:NOgrad}
\end{figure}

\begin{figure}  
\centering
\includegraphics[width=\columnwidth]{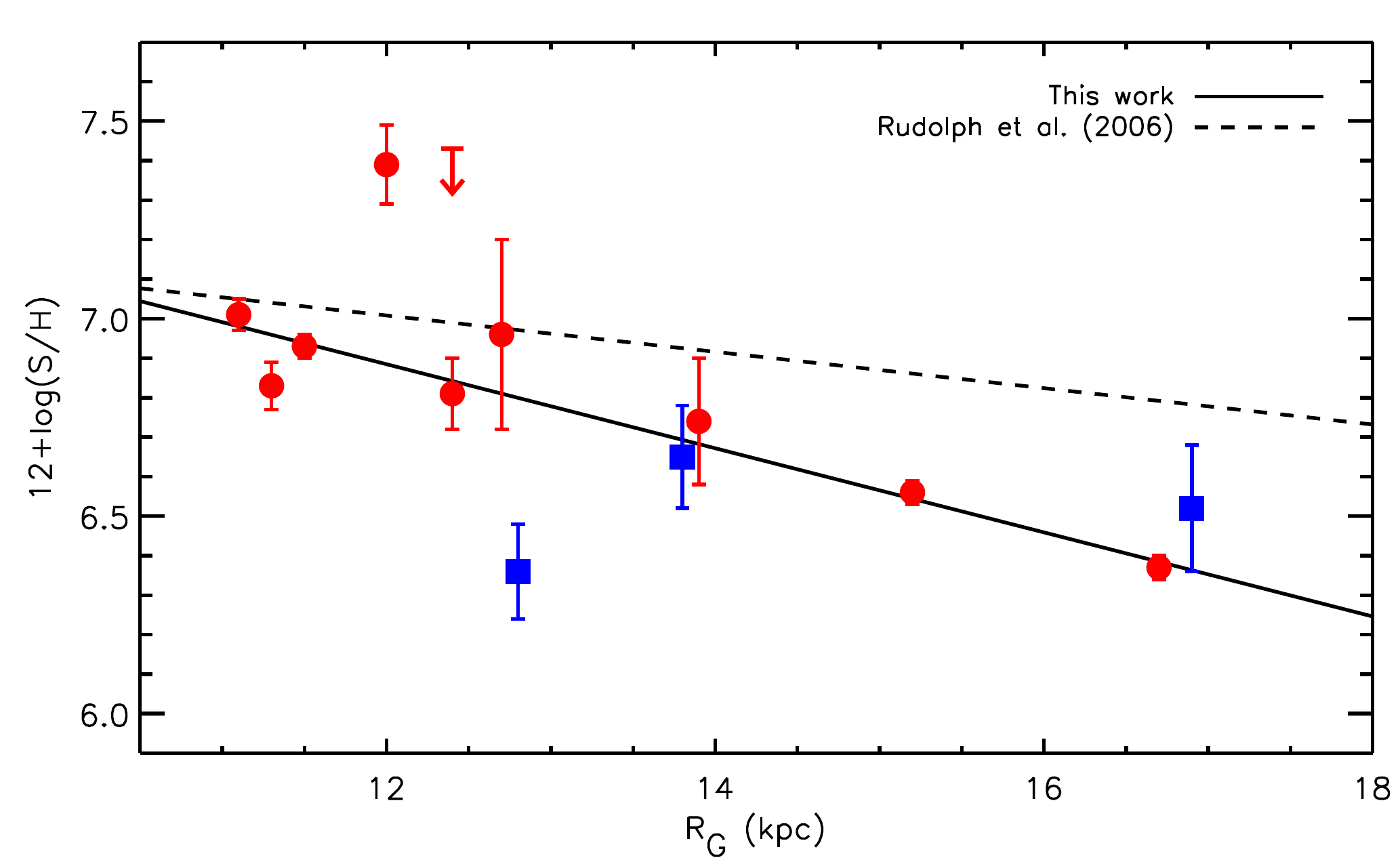}
\caption{Radial distribution of 12+log(S/H) plotted versus Galactocentric radius. Symbols and colours are as in Fig. \ref{fig:OHgrad}.}
\label{fig:SHgrad}
\end{figure}

\begin{figure}  
\centering
\includegraphics[width=\columnwidth]{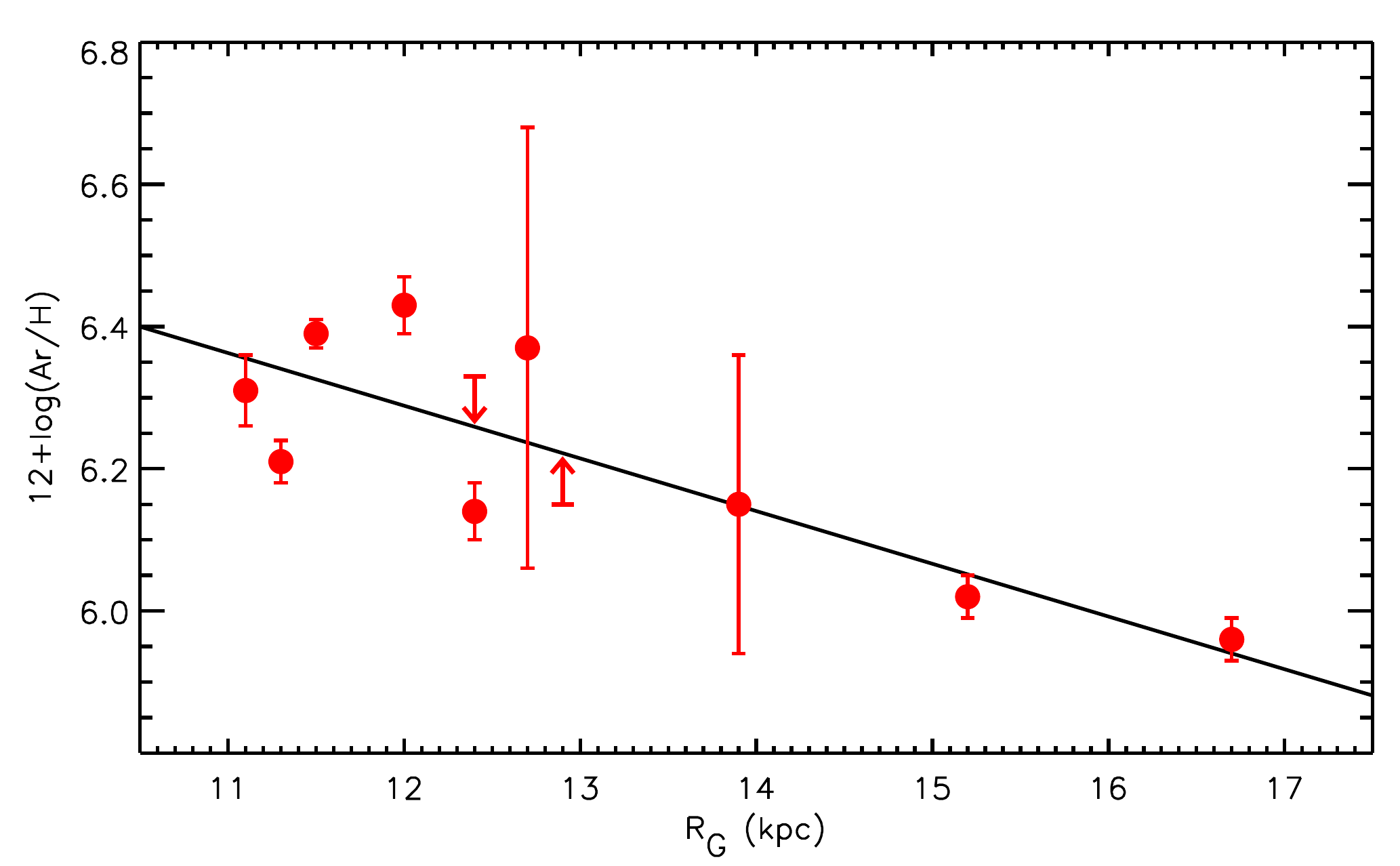}
\caption{Radial distribution of 12+log(Ar/H) plotted versus Galactocentric radius. Symbols and colours are as in Fig. \ref{fig:OHgrad}.}
\label{fig:ArHgrad}
\end{figure}

\begin{figure}  
\centering
\includegraphics[width=\columnwidth]{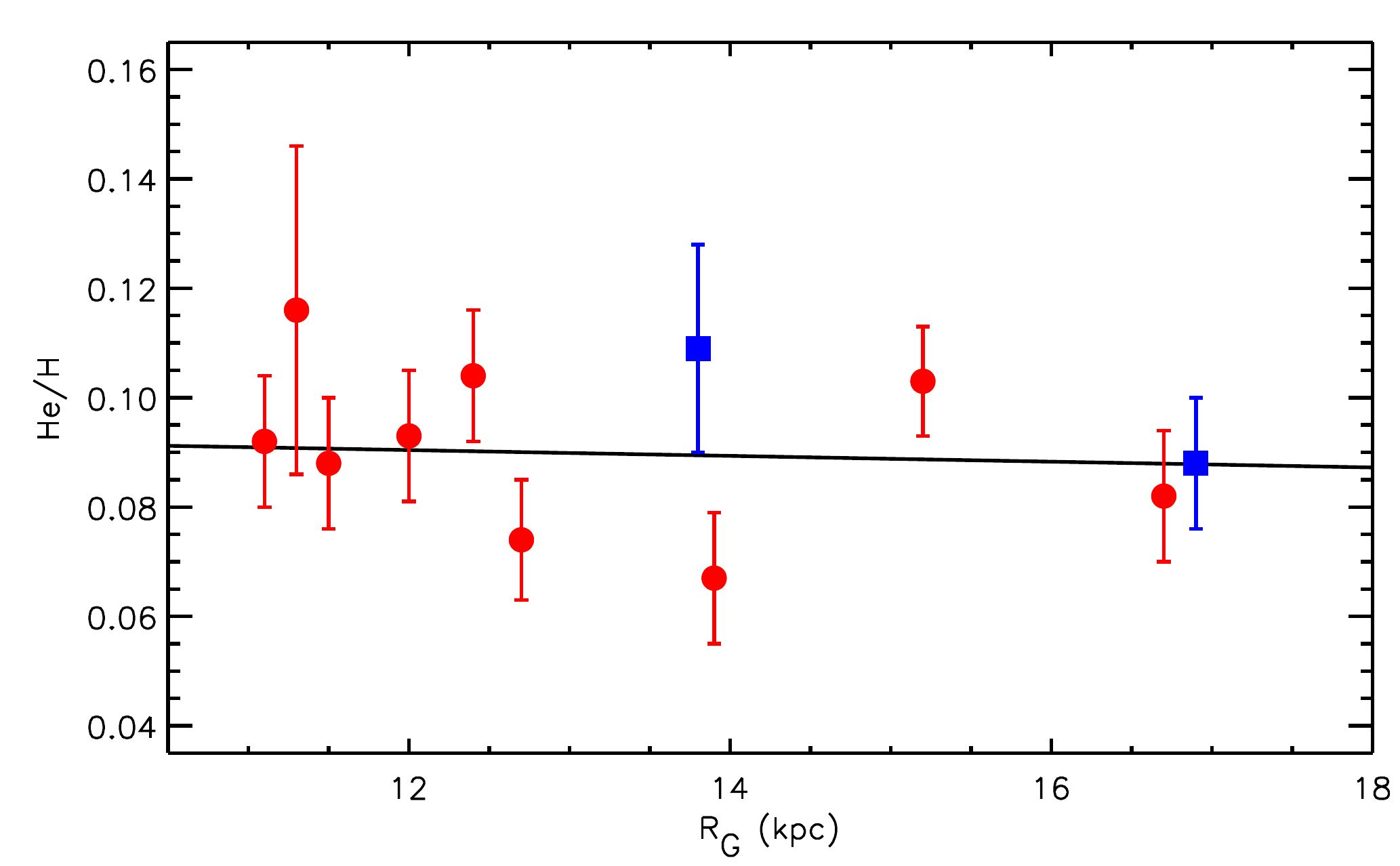}
\caption{Radial distribution of He/H plotted versus Galactocentric radius. Symbols and colours are as in Fig. \ref{fig:OHgrad}.}
\label{fig:HeHgrad}
\end{figure}

From the derived abundances shown in Tables \ref{table:abundances}, \ref{table:models}, and \ref{table:Tr_abundances}, we can study the radial distribution of chemical abundances 
along the Galactic anticentre. Figures \ref{fig:OHgrad} through \ref{fig:HeHgrad} show the total chemical abundances of O/H, N/H, N/O, S/H, Ar/H, and He/H, respectively, 
plotted against Galactocentric distances for the \HII~regions with chemical abundances inferred in this work. To make it easier to distinguish between the three methods 
used to determine abundances, the regions with direct T$_e$ are plotted as red circles, results from tailor-made models are represented with black stars, and abundances 
obtained from the t$_e$-t$_e^*$ relation are shown as blue squares. Regions with abundances that are upper or lower limits are represented by arrows in the figures, but 
they have not been taken into account in any fitting. 

To determine linear gradients we performed a least-squares fitting weighted by abundance uncertainties. Data from the three methods (except limit abundances) were combined 
into a single fit and plotted as solid lines in the figures. We also computed gradients using only abundances derived for the \HII~regions with direct determinations of T$_e$; 
however, for simplicity, they are not represented in the figures. The final abundance gradients are presented in Table \ref{table:gradientes}. Regions with chemical 
abundances derived with two methods (S98 and S228) are represented twice in the figures, but are only consider once in the performed fits; as the representative value we 
took the weighted mean by errors of the two methods. In addition, it is interesting to note that those abundances estimated using the electron temperature derived from the 
relation between optical and radio temperatures (blue squares) follow the same tendency as abundances derived with the direct method and, in general, they agree with the 
fits obtained.

To compare our results with other works, in Table \ref{table:grad_literature} we present a summary of all the gradients obtained by different authors from \HII~regions in 
the inner and outer Galaxy in several spectral ranges. The extrapolated radial gradients obtained by \citet{Rudolph2006} with optical data are also represented in graphics 
as dashed lines. These come from a selection of previous works for a large range of Galactocentric distances shown here for the sake of comparison with a suggestion of 
flattening in the outer parts of the disc.\\

The radial distribution of the oxygen abundance is presented in Fig. \ref{fig:OHgrad} and it clearly shows a decrease with Galactocentric distance. 
Within the range of distances between 11 and 18 kpc, the maximum variation goes from 12+log(O/H)=8.44 to 12+log(O/H)=8.05. The results obtained by computing the fit with 
all the regions and by using oly the abundances derived with the direct method (Table \ref{table:abundances}) are consistent, taking the associated uncertainties into 
account; the slope of the latter is slightly steeper.

When comparing our results with those for the inner parts of the Galaxy, we find that our slope is clearly shallower than the one obtained 
by \citet{Shaver1983}, but in general their fits have steeper slopes than many others in the literature. In addition, our result is in very good agreement with the gradient 
obtained by \citet{Afflerbach1997} with infrared data. The slope obtained by \citet{Esteban2005} is smoother than ours, even though they performed the study of gradients based 
on recombination lines instead of collisional lines. On the other hand, our gradient is in good agreement with those performed in a wider range of distances (0 to18 kpc) 
and, as can be seen in Fig. \ref{fig:OHgrad}, our slope agrees with \citet{Rudolph2006}. Gradients obtained by \citet{Deharveng2000}, \citet{Quireza2006}, and \citet{Balser2011} 
are slightly shallower than ours, but are consistent within errors. \citet{Vilchez1996} are the only authors who have focused their study on the outer part of the Galaxy. They 
provide three different oxygen gradients, and our slope agrees with the one obtained from all of their data (third option in Table \ref{table:grad_literature}), but not 
with the other fits.\\

Figure \ref{fig:NHgrad} shows the radial distribution of the final N/H abundances. The derived nitrogen abundances present values between 12+log(N/H)=7.44 and 6.88 and their 
distribution shows a clear decrease with distance. Observed values follow closely the performed fit, although those derived from radio temperatures (blue squares) are 
slightly higher, but still in agreement within measured errors. The resulting fits, for all the regions and for direct abundances alone, are in agreement, taking the errors 
into account.

Within the uncertainties this result for the nitrogen gradient is not different from those found by \citet{Shaver1983}, \citet{Simpson1995}, \citet{Rudolph1997}, 
\citet{Afflerbach1997}, and \citet{Rudolph2006}. However, our slope differs from that obtained by \citet{Vilchez1996} for a similar range of R$_G$: we do not see any 
clear steepening of nitrogen in the outer part of the Galaxy disc.\\

Figure \ref{fig:NOgrad} shows the behaviour of N/O versus Galactocentric distance for the studied regions. The derived values of N/O range from log(N/O)=-0.87 to -1.19. In this 
case almost no difference exists between the slope derived from directly derived abundances and the value considering also abundances derived using model fits, since the 
latter have much higher associated errors and they thus have a lower weight in the fit. Our slope is consistent with the value obtained by \citet{Simpson1995} from infrared 
observation in the inner parts of the Galaxy, but differs from \citet{Rudolph2006} who found that N/O is nearly constant at all distances. The fit of the data shown in 
Fig. \ref{fig:NOgrad} presents a decreasing slope in N/O with Galactocentric distance, although a step function is also plausible (see next section). \\

The sulphur radial gradient is plotted in Fig. \ref{fig:SHgrad}. The derived sulfur abundances present values between 12+log(S/H)=7.01 and 6.37 along the range of 
Galactocentric distances, with the exception of the high S/H value of S311 (which has the lowest value of t$_e$(\SIII)) and the low estimation of S/H in S98. The distribution 
shows a negative radial gradient with a slope that is much steeper than those derived by other authors. This may be a consequence of the large weight of regions located 
at large Galactic radius whose abundances have very low errors.\\

In Fig. \ref{fig:ArHgrad}, we present the argon gradient along the Galactocentric anticentre. The distribution ranges from 12+log(Ar/H)=6.4 to 5.9, decreasing clearly with 
distance.  In this case all the abundances represented were derived using the direct method (direct measures of electron temperature for all the regions) and we only show 
one fit for the gradient in Table \ref{table:gradientes}. This is the first time that the distribution of argon towards the Galactic anticentre is shown. \citet{Shaver1983} 
derived a gradient of d(log(Ar/H))/dR$_G$=-0.06$\pm$0.015 in the inner part of the Galaxy, which is in good agreement with our slope, taking the fit uncertainties into account.\\

Finally, Fig. \ref{fig:HeHgrad} shows the radial distribution of helium abundances. The derived values of He/H cluster around He/H$\sim$0.09 no matter their Galactocentric 
distances. There is no obvious gradient and the least-squares fit to the data gives a slope that is not significantly different from zero. Trying to compare our results 
with those of other authors, we find that no attempts have been made previously to estimate the total helium radial abundance gradient of the Galaxy, only the distribution 
of He$^+$/ H$^+$ has been studied before with slopes also close to zero \citep{Peimbert1978,Talent1979,Shaver1983,Deharveng2000}.

\subsection{Chemical composition of the outer Galaxy}\label{sect:distribution}
\begin{figure}  
\centering
\includegraphics[width=\columnwidth]{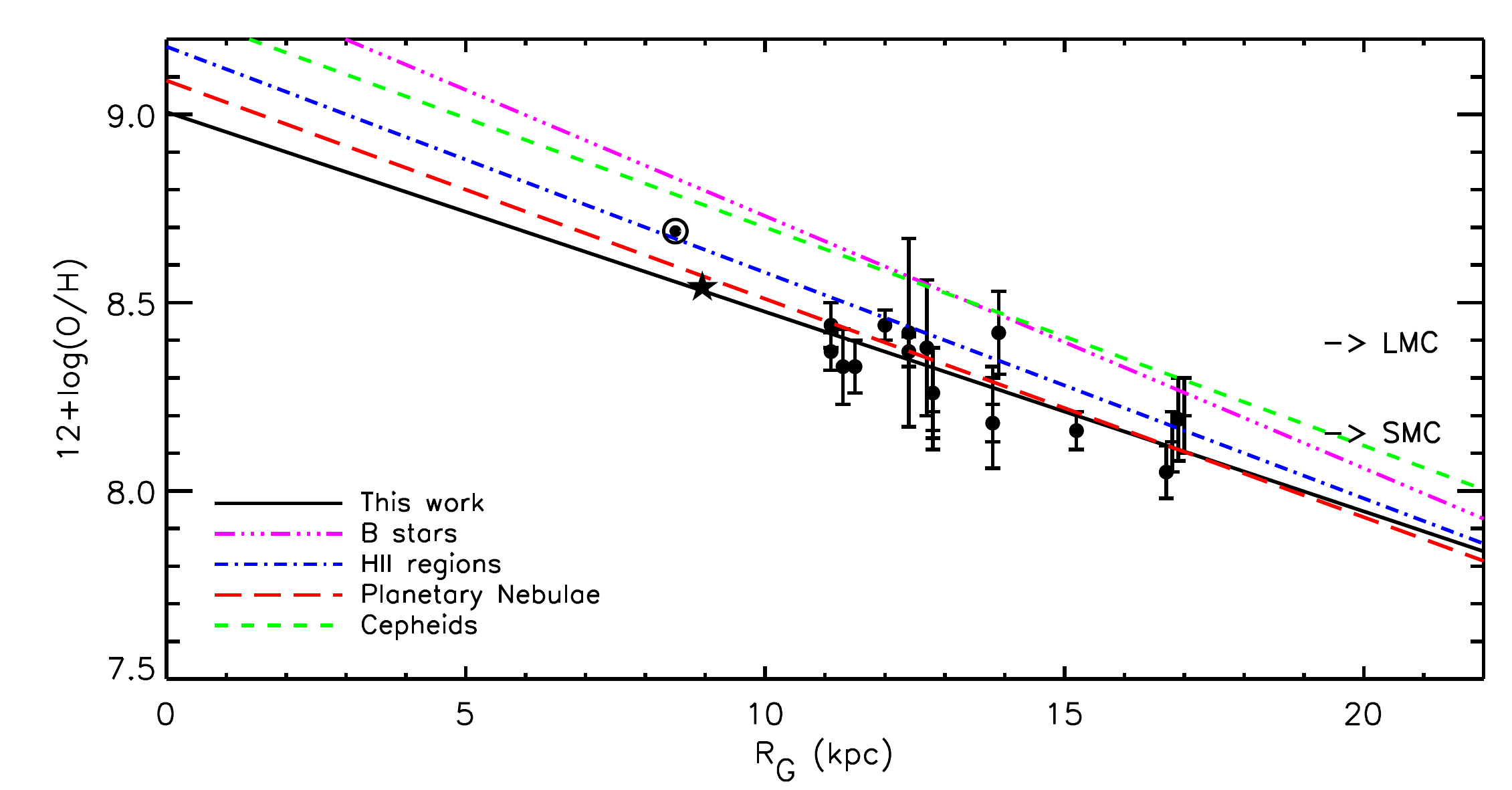}
\caption{Extrapolated gradients of 12+log(O/H) along the Galactic disc for different sources. Black circles represent all the \HII~regions studied in this work and the 
solid black line indicates the gradient obtained with them. Gradients derived from PNs (long dashed red line), \HII~regions (dash-dotted blue line), B-type stars 
(dash-dot-dotted pink line) and Cepheids (short-dashed green line) are also represented. The Sun and the Orion nebula (M42) are shown at their respective O/H abundances 
located at R$_G$ =8.5 kpc and R$_G$=8.95 kpc, respectively. Horizontal arrows show chemical abundances of the Small and Large Magellanic Clouds.}
\label{fig:gradOHgrande}
\end{figure}

The knowledge of the chemical distribution across the Galactic disc is a benchmark for models of chemical evolution. For this reason, considerable effort has been made to 
establish the abundance gradients in the MW by studying the distribution of many elements and many sources which all probe different epochs in the MW chemical 
evolution history. Figure \ref{fig:gradOHgrande} provides comparisons of our O/H abundance gradient derived from the total \HII~region sample towards the anticentre with  
many other gradients derived from studies with PNe, cepheids, B-type stars, and \HII~regions in a wide range of Galactic distances (0-20 kpc). The gradients adopted 
for the different objects are shown in Table \ref{table:grad_sources} and are those provided directly by the authors. Moreover, oxygen abundances derived in the Large and 
Small Magellanic Clouds \citep{Russell1990}, Orion nebula \citep{Tsamis2011}, and the Sun \citep{Asplund2009} are also represented in Fig. \ref{fig:gradOHgrande}.

B-type stars are massive young stars ($<$10~Myr) and their abundances trace the metallicity near their current location. \citet{Rolleston2000} studied a sample of about 80 
early B-type main sequence stars located from 6 to 18 kpc. They conclude that the distribution of oxygen can be represented by a linear gradient 
with no evidence of a better fitting with a two-zone model. As we can see in Fig. \ref{fig:gradOHgrande} and Table \ref{table:grad_sources}, the results 
from \citet{Rolleston2000} are the most different to ours, with a much steeper slope and the largest ordinate at origin.

Planetary nebulae trace metallicity in the Galactic disc, bulge, and halo. Their ages span 1-8 Gyr and thus PNe can in principle probe metallicity with time. We compare 
our results with those obtained from \citet{Henry2010} who determined chemical abundances of 124 PNe extending from 0.9 to 21 kpc in Galactocentric distances. We find 
an agreement in slope with their gradient as we can see in Fig. \ref{fig:gradOHgrande} and Table \ref{table:grad_sources}. They found some evidence that the gradient 
beyond 10 kpc is steeper than it is inside this distance.

Cepheids are evolved stars that can be detected at large distances and are relatively young ($<$200Myrs), hence they are primary tracers of metals in the Galactic disc. 
We compare our gradient with that obtained by the recent work of \citet{Korotin2014} and find a good agreement between the two slopes. They also found some evidence that 
the distribution might become flatter in the outer parts of the disc.

Galactic \HII~regions are the formation sites of massive OB stars and they reveal the location of current Galactic star formation. In Fig. \ref{fig:gradOHgrande} we also 
represent the gradient obtained by \citet{Rudolph2006}, considering it as a collection of previous studies with \HII~regions. Their results are consistent with ours, 
taking the uncertainties into account.\\

\begin{table}
\caption{Comparison of oxygen gradient obtained from different sources.} 
\label{table:grad_sources} 
\centering 
\begin{tabular} {l  c c c } 
\\
\hline
Author(s)	 & Object  &   d(log(O/H))/dR$_G$ & 	$\Delta$R$_G$  \\	
	 & type  &   (dex/kpc$^{-1}$) & 	(kpc)  \\	
\hline\hline
This work  &  \HII~regions & -0.053  & 11-18\\
\citet{Rolleston2000}  &  B stars   &   -0.067   & 6-18 \\
\citet{Rudolph2006} & \HII~regions &  -0.060 & 5-18 \\
\citet{Henry2010}  & PNe  & -0.058  & 0.9-21\\
\citet{Korotin2014}   & Cepheids  & -0.058  & 5-18\\
\hline
\end{tabular}
\end{table}

What do we conclude from these studies?  The existence of those gradients offers the opportunity to compare with models of disc evolution to understand the formation 
and chemical evolution of our Galaxy. All the results obtained from different sources present a negative radial gradient (although with significant differences in their 
slopes), thus metal abundances in the inner disc are higher than those in the outer disc. This result supports the Inside-Out formation scheme of the 
disc \citep{Hou2000,Alibes2001,Chiappini2003}. The inner part of the disc always forms first, and the outer part forms progressively later as gas with higher angular 
momentum settles into the equatorial plane at larger radii, causing the disc to grow outward with time. The metal abundance of the innermost part of the disc is higher 
because of the prior enrichment of the infalling gas before it reaches the equatorial plane, and the metal abundances of the outermost part of the disc is smaller because 
of the relatively long time scale for star formation and metal enrichment. Therefore, the negative gradients obtained in this work and results from other sources are consistent 
with Inside-Out models; the differences in slopes can be explained by the amount of material, yields, and star formation rates adopted by the model (e.g. \citealt{Molla2015}).\\

As far as the shape of the gradient is concerned, the possibility of variations in the slope towards the outer disc is still widely debated. Several works 
(not represented in Fig. \ref{fig:gradOHgrande} for clarity) based on different tracers (cepheids, open clusters, PNe, and \HII~regions) have found evidence 
that the radial abundance gradient of some elements may flatten out at the outer part of the Galaxy disc 
\citep{Fich1991,Vilchez1996,Maciel1999,Andrievsky2004, Costa2004, Luck2006,Carraro2007, Pedicelli2009,Henry2010,Andreuzzi2011,Korotin2014}. 
In addition, strong evidence for flattening at large distances of discs of other spiral galaxies has been presented in several papers 
\citep{vanZee1998,Goddard2011,Werk2011,Bressolin2012}.

Our sample is located towards the anticentre; in order to analyse the radial behaviour of the abundance gradient in the outermost disc, we need to make a comparison 
with literature gradients performed in wider radial ranges including the inner parts of the Galaxy. When doing so, our data suggest a shallower slope. Although this result 
does not necessarily imply a flattening of the gradient in the outer disc, what is clear is that the simple extrapolation of the inner slope towards the anticentre is not 
consistent with the observations.

\begin{figure}[t!]  
\centering
\includegraphics[width=\columnwidth]{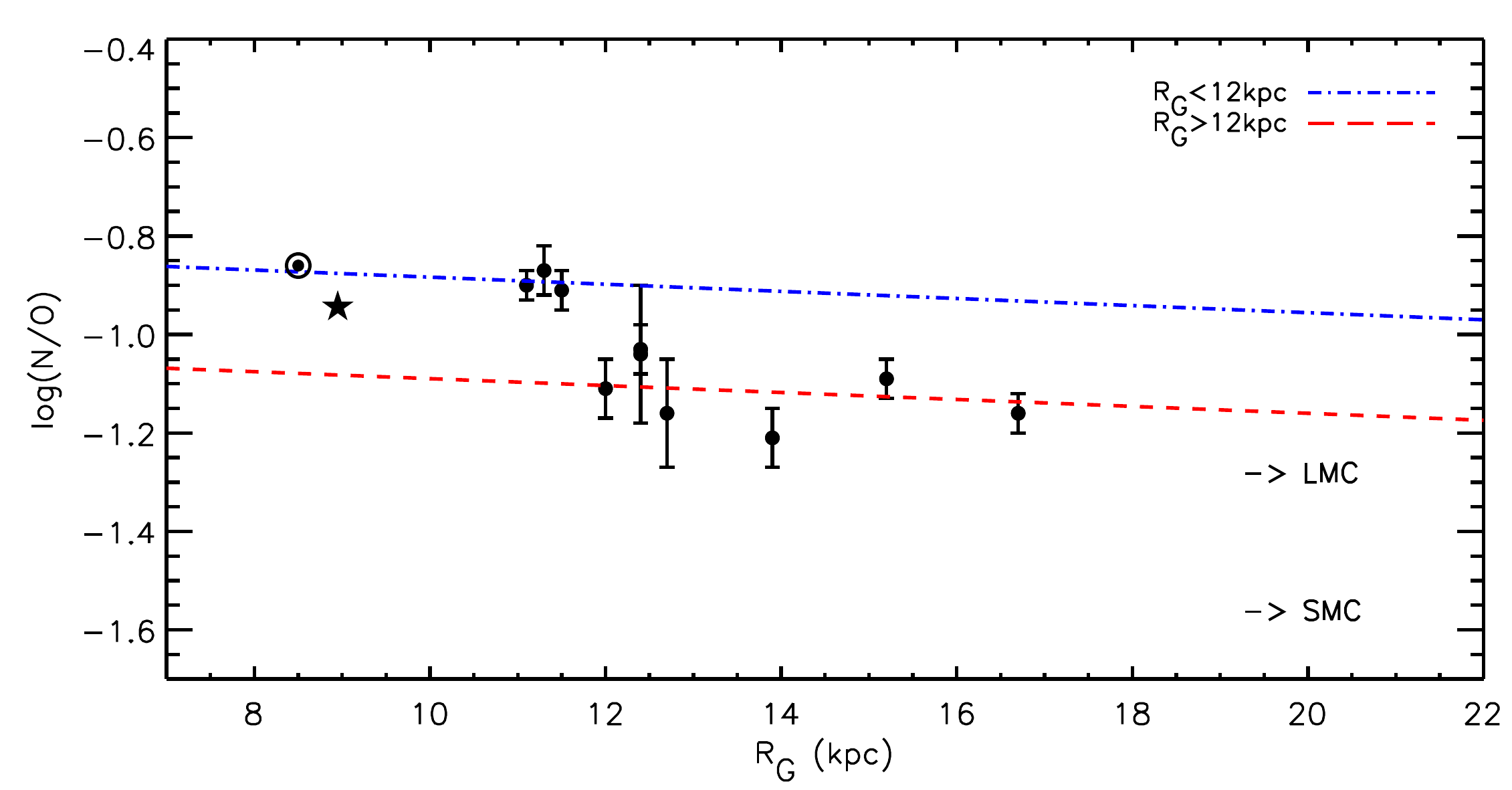}
\caption{Radial distribution of N/O plotted versus Galactocentric radius. Black points represent N/O abundances obtained with the direct method. The two lines represent 
the fit to data in two distances ranges: within 12 kpc from the Galactic centre (blue dot-dashed line) and farther than this limit (red dashed line), see text for details. 
The Orion nebula, the Sun and Magellanic Clouds are represented as in Fig. \ref{fig:gradOHgrande}.}
\label{fig:gradNOgrande}
\end{figure}

From the theoretical point of view, several chemical evolution models have been computed to reproduce the flattening observed. However, authors suggest different physical 
mechanisms to explain this behaviour: lower effective yields in the outer disc than in the inner disc \citep{Twarog1997}, effective suppression of mixing processes near the 
corotation circle \citep{Andrievsky2002,Luck2003,Andrievsky2004}, constant density distribution of the halo stellar component in the inner 20 kpc \citep{Cescutti2007}, 
levelling out of the star formation efficiency about and beyond the isophotal radius \citep{Esteban2013}, etc. More information about this topic can be found 
in \citet{Bressolin2012}, who discuss several mechanisms that can be responsible for the flattening in the outermost disc of spiral galaxies.  \\

The N/O abundance ratio is a useful indicator of the chemical age of the galaxy and how evolved the disc is. An interesting feature suggested by Fig. \ref{fig:NOgrad} is that 
our N/O gradient seems to be better represented by a step function instead of a linear fit. We compute a new double fit using only those N/O abundances derived from the direct 
method (Table \ref{table:abundances}) and differentiating those regions placed within 12 kpc from the centre and farther than this limit. The result for R$_G<$12~kpc is 

log(N/O) = - (0.7264 $\pm$ 1.3963) + (0.0152 $\pm$ 0.1240)~R$_G$,

while for R$_G>$12kpc we obtain

log(N/O) = - (0.8993 $\pm$ 0.1689) + (0.0150 $\pm$ 0.0116)~R$_G$.

Figure \ref{fig:gradNOgrande} shows the radial distribution of direct N/O abundances within the two fits performed with data at R$_G<$12 kpc (blue dot-dashed line) 
and R$_G>$12 kpc (red dashed line). As can be seen, both data sets are represented by lines with similar slopes (nearly constant) but different origins as the N/O ratio 
is uniform in two distance ranges. We conclude that the N/O distribution with Galactic distance for the disc can be better described by a step function rather than a 
linear gradient. Furthermore, with this double fit the statistical dispersion is obviously smaller than with the single fit.

\begin{figure}  
\centering
\includegraphics[width=\columnwidth]{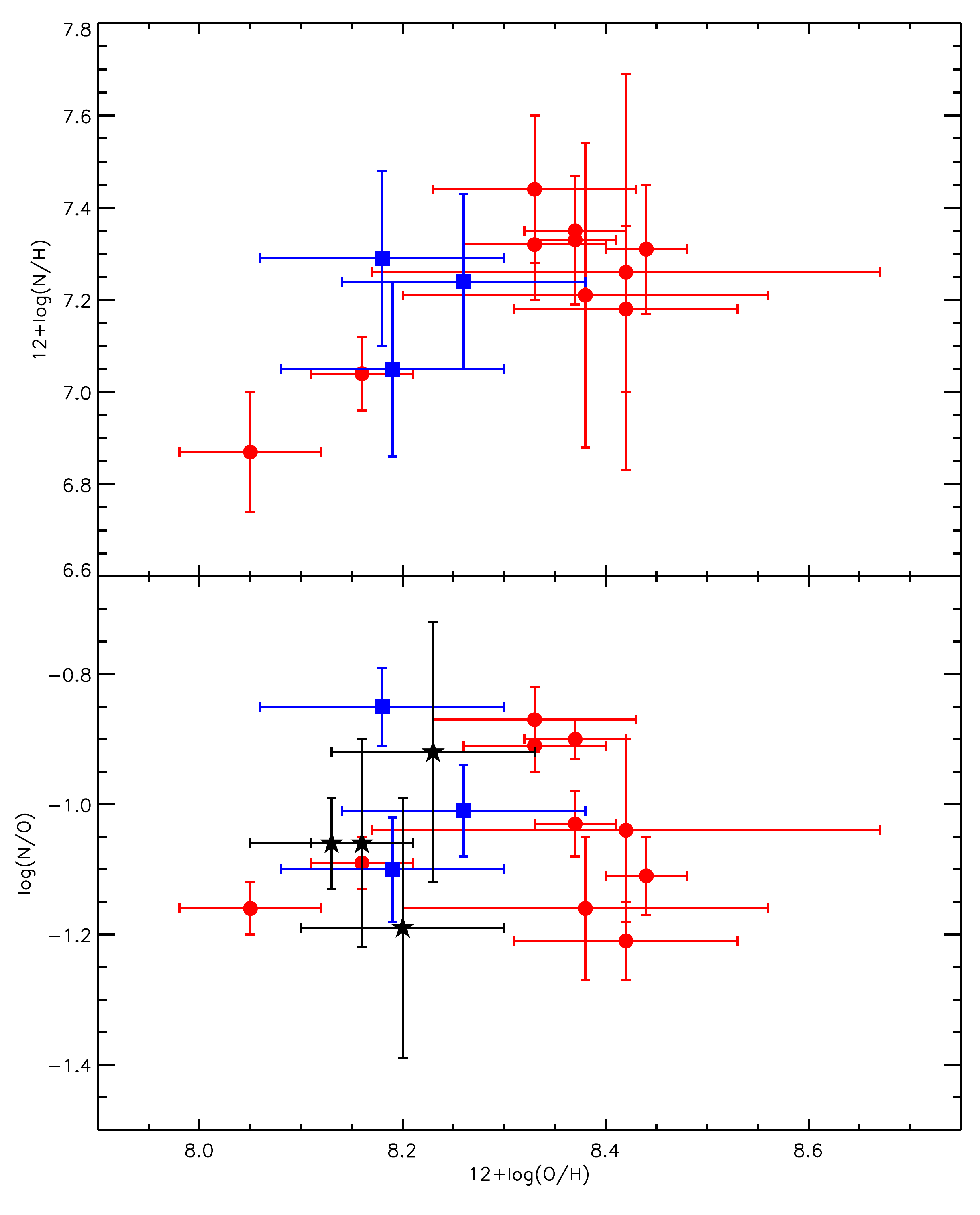}
\caption{Top panel: Relation between nitrogen and oxygen abundances. Bottom panel: Relative abundance, log(N/O), versus the oxygen abundance. Symbols and colours are as in 
Fig. \ref{fig:OHgrad}.}
\label{fig:compNO}
\end{figure}

From the point of view of chemical evolution, the origin of nitrogen is still a matter of important debate because it can have a primary or secondary origin. The overall 
picture of the chemical composition of galaxies seems to indicate that there are two main behaviours in the N/O versus O/H relationship depending on metallicity: a flat 
line with log(N/O)=-1.46 \citep{Garnett1990} for the data of low-mass and irregular galaxies with 12+log(O/H)$<$8.3, and a strong slope for the N/O data against oxygen 
abundance for metal-rich regions.  The first case reflects the proportion of the true primary production of nitrogen (independent of oxygen abundance), which is believed 
to be synthesised in massive stars. The second situation explains the existence of secondary nitrogen (proportional to oxygen abundance) processed in the CNO burning during 
the stellar evolution and returned into the ISM via PNe or stellar winds of massive stars (e.g. \citealt{Edmunds1990}, \citealt{Molla2006}).

In Fig. \ref{fig:compNO} we analyse the behaviour of nitrogen abundances (top panel) and N/O ratio (bottom panel) versus oxygen abundances for the regions located towards
the anticentre studied in this work. As can be seen, nitrogen follows oxygen and the N/O ratio is constant with O/H, suggesting that nitrogen has a primary origin. 
However, our data show that in this zone log(N/O)$>$-1.46, which is indicative of a previous enrichment by evolved 
intermediate mass stars (the secondary nitrogen component). In this case, the fact that N/O is constant informs us that all regions have a similar 
\textquotedblleft chemical age\textquotedblright (understanding age in the sense of the time since the bulk of its star formation occurred). The observed N/O values and radial
 behaviour in the outer parts of the disc compared to the solar neighbourhood can be understood if we are witnessing a special time in the chemical evolution of the outer 
MW; in the \citet{Edmunds1978} chemical age scenario this can be a result of the increase in nitrogen abundance from intermediate-mass stars. When comparing our nitrogen 
distribution with the abundances in the Magellanic Clouds, it can be seen that our N/O ratio appears larger than the value for the SMC and LMC, log(N/O)= -1.58 $\pm$ 0.16 
and log(N/O)= -1.30 $\pm$ 0.30, respectively \citep{Russell1990}, for a range in log(O/H) comparable to the abundances measured in the outer disc; this is in line with the 
chemical ageing scenario mentioned before for the outer disc of the MW.

%SECTION 5
%%%%%%%%%%%%%%%%%%%%%%%%%%%%%%%%%%%%%%%%%%%%%%%%%%%%%%%%%%%%%%%%%%%%%%%%%%%%%%

\section{Summary and conclusions}\label{sect:conclusions}
The abundances of heavy elements in the ISM provide a fossil record of the enrichment which has taken place due to the nucleosynthesis in successive generation stars. The 
study of the radial variations of metallicity across the Galactic disc is a powerful method for understanding the history of star formation and chemical evolution of the MW. 
The primary focus of this paper was to give an accurate description of the radial chemical distribution towards the Galactic anticentre up to a distance of 18~kpc.

We have carried out new optical spectroscopic observations of nine \HII~regions located towards the Galactic anticentre using the ISIS spectrograph at the WHT. To realise a 
more extensive study, the sample was increased by searching for optical \HII~regions at R$_G>$11~kpc in the literature. The complete sample comprised 23 \HII~regions extending 
in Galactocentric radius from 11~kpc to 18~kpc.

For 13 regions of the sample we have derived accurate electron densities from the \SII $\lambda\lambda$6717,6731 line ratio and electron temperatures 
t$_e$(\OIII), t$_e$(\SIII), t$_e$(\NII), t$_e$(\OII), and t$_e$(\SII). These physical parameters have been used to determine direct total chemical abundances. We have also 
performed tailor-made photoionisation models to derive chemical abundances for those regions without direct estimations of temperature. In addition, by comparing radio 
recombination temperatures from the literature and optical temperatures from this work, we have obtained the empirical relation 
t$_e$(\OIII)=(0.175$\pm$0.080)+(0.792$\pm$0.081)$\times$t$_e^*$. This relation has been applied to three regions without direct temperature to obtain their physical parameters 
and abundances.\\

To study the radial gradients of metallicity across the outermost part of the MW, we have performed weighted least-squares fits to the distribution of the O/H, N/H, N/O, 
S/H, Ar/H, and He/H abundances along the Galactocentric distances.  We have concluded the following:

\begin{enumerate}

\item The radial distribution of oxygen abundance is fit well by a regression whose form is 12+log(O/H)=(9.006$\pm$0.112)-(0.053$\pm$0.009)R$_G$, where R$_G$ is the 
Galactocentric radius in kpc. When comparing our result with gradients performed with several objects in wider radial ranges, including the inner parts of the Galaxy, 
we find a suggestion for a shallower slope. Our data cannot be used to confirm the flattening of the abundance gradient towards the outer disc suggested previously, but 
clearly show that a linear extrapolation of the inner slope is inconsistent with the observations

\item The derived gradient for N/H abundances shows a clear decrease with distance with a slope of -0.080$\pm$0.019 dex~kpc$^{-1}$. This result is in good agreement with 
previous works. We do not see any steepening of nitrogen in the outer part of Galaxy disc.

\item In the case of the N/O ratio, the radial distribution is better described by a step function rather than a linear gradient. The double-fit performed shows different 
behaviours for regions located beyond 12~kpc than for those with R$_G<$12~kpc, but both with constant slopes. The radial behaviour in the outer parts of the disc compared 
to the solar neighbourhood can be understood if we are witnessing a special time in the chemical evolution of the outer MW, which can be a result of the increase in nitrogen 
abundance from intermediate-mass stars.

\item The fit obtained for the distribution of sulphur abundance is 12+log(S/H)=(8.162$\pm$0.088)-(0.106$\pm$0.006)R$_G$. This gradient shows a slope that is much steeper 
than those derived by other authors, probably reflecting the more accurate values from our data at large Galactic distances.

\item The argon gradient has been derived for the first time in regions towards the Galactic anticentre with the functional 
form 12+log(Ar/H)=(7.178$\pm$0.073)-(0.074$\pm$0.006)R$_G$. This result is in good agreement with the value obtained in the inner part of the disc.

\item The total helium radial abundance gradient has been also presented here for the first time. The least-squares fit to the data gives a slope that is not 
significantly different from zero. 

\end{enumerate}

\begin{acknowledgements}
This work has been partially funded by the project AYA2013-47742-C4-1. A.M. acknowledges funding from the Spanish AYA2007-66804 and AYA2012-35330 grants. We thank 
D. D\'iaz-Fraile, A. Sota, J. Blasco, and the ESTALLIDOS collaboration for their useful comments and scientific support.
\end{acknowledgements}

%TABLES

\onecolumn
\begin{landscape}
\small
 \setlength\tabcolsep{2pt}
 \begin{center}
\setcounter{table}{2} 
\begin{longtable}{l c c c c c c c c c c c}
\caption{Reddening corrected line intensities relative to \Hb=1000 of the \HII~regions observed with ISIS-WHT.} \tabularnewline \hline
\label{table:sample_intensities}
	&&&\multicolumn{9}{c}{I($\lambda$)/I(\Hb)} \tabularnewline
	 \cline{4-12} \tabularnewline
	Line &   $\lambda$~(\A)&   f($\lambda$)&    S83&   S132&    S156&    S162& S207&    S208&    S212&    S228&    S270 \\
	\hline \hline   \\
	\endfirsthead
	\caption{Continued.} \\
	\hline
	&&&\multicolumn{9}{c}{I($\lambda$)/I(\Hb)} \\
	 \cline{4-12}
	Line &   $\lambda$~(\A)&   f($\lambda$)&    S83&   S132&    S156&    S162& S207&    S208&    S212&    S228&    S270 \\
	\hline \hline   \\
	\endhead
	\\
 \hline	
  Continues. \\
	 \endfoot
	 \endlastfoot
\OII	&	3727.43	&	0.322	&	908.8	$\pm$ 	87.0		&	4119.4	$\pm$ 	393.2		&	3015.6	$\pm$ 	244.0		&	2572.5	$\pm$ 	85.3		&	4497.0	$\pm$ 	331.0		&	3375.0	$\pm$ 	270.4		&	2959.8	$\pm$ 	204.3		&	3673.2	$\pm$ 	311.8		&	2328.3	$\pm$ 	114.1	$\dagger$ 	\\
H13	&	3734.37	&	0.320	&	-		 		&	-		 		&	-		 		&	11.1	$\pm$ 	2.0		&	-		 		&	-		 		&	-		 		&	-		 		&	-		 		\\
H12	&	3750.15	&	0.317	&	-		 		&	42.1	$\pm$ 	8.6		&	37.5	$\pm$ 	3.2		&	32.6	$\pm$ 	1.8		&	-		 		&	-		 		&	26.8	$\pm$ 	2.8	$\dagger$	&	-		 		&	-		 		\\
H11	&	3770.63	&	0.313	&	-		 		&	51.7	$\pm$ 	8.1		&	42.5	$\pm$ 	3.8		&	36.6	$\pm$ 	1.3		&	-		 		&	-		 		&	37.4	$\pm$ 	4.7	$\dagger$	&	-		 		&	-		 		\\
H10	&	3797.90	&	0.307	&	-		 		&	63.3	$\pm$ 	7.3		&	58.3	$\pm$ 	4.8		&	51.7	$\pm$ 	1.7		&	-		 		&	-		 		&	39.4	$\pm$ 	4.0	$\dagger$	&	-		 		&	-		 		\\
\HeI	&	3819.61	&	0.302	&	-		 		&	-		 		&	7.0	$\pm$ 	0.9		&	11.9	$\pm$ 	0.8		&	-		 		&	-		 		&	-		 		&	-		 		&	-		 		\\
H9+HeI	&	3834.48	&	0.299	&	-		 		&	86.6	$\pm$ 	9.8		&	82.2	$\pm$ 	6.4		&	82.8	$\pm$ 	2.3		&	-		 		&	-		 		&	94.8	$\pm$ 	9.3		&	-		 		&	-		 		\\
\NeIII	&	3869.35	&	0.291	&	560.4	$\pm$ 	64.2		&	-		 		&	13.9	$\pm$ 	1.5		&	26.3	$\pm$ 	1.2		&	-		 		&	-		 		&	-		 		&	-		 		&	-		 		\\
H8+\HeI	&	3889.05	&	0.286	&	284.9	$\pm$ 	26.5		&	269.9	$\pm$ 	28.0		&	202.6	$\pm$ 	14.6		&	191.2	$\pm$ 	5.2		&	248.4	$\pm$ 	31.8	$\dagger$	&	-		 		&	274.3	$\pm$ 	23.3		&	256.3	$\pm$ 	30.5	$\dagger$ 	&	-		 		\\
H7+\NeIII	&	3968.77	&	0.266	&	227.7	$\pm$ 	43.2		&	238.9	$\pm$ 	25.5		&	205.5	$\pm$ 	13.8		&	192.6	$\pm$ 	4.9		&	206.5	$\pm$ 	27.5	$\dagger$	&	-		 		&	234.4	$\pm$ 	16.4		&	157.5	$\pm$ 	20.9	$\dagger$ 	&	-		 		\\
\HeI	&	4026.21	&	0.251	&	-		 		&	-		 		&	18.7	$\pm$ 	1.5		&	20.1	$\pm$ 	0.8		&	-		 		&	-		 		&	-		 		&	-		 		&	-		 		\\
\SII	&	4068.60	&	0.239	&	-		 		&	41.4	$\pm$ 	4.6		&	23.6	$\pm$ 	1.6		&	24.7	$\pm$ 	0.9		&	-		 		&	-		 		&	11.5	$\pm$ 	1.1	$\dagger$ 	&	-		 		&	-		 		\\
\SII	&	4076.35	&	0.236	&	-		 		&	23.0	$\pm$ 	5.6	$\dagger$ 	&	8.6	$\pm$ 	0.9		&	8.6	$\pm$ 	0.6		&	-		 		&	-		 		&	-		 		&	-		 		&	-		 		\\
\Hd	&	4101.74	&	0.229	&	310.1	$\pm$ 	36.1		&	322.2	$\pm$ 	22.9		&	313.1	$\pm$ 	18.1		&	276.9	$\pm$ 	5.9		&	352.3	$\pm$ 	32.7	$\dagger$	&	414.3	$\pm$ 	50.7	$\dagger$	&	305.7	$\pm$ 	15.4		&	312.0	$\pm$ 	28.7		&	-		 		\\
\Hg	&	4340.47	&	0.156	&	535.4	$\pm$ 	28.6		&	523.7	$\pm$ 	24.6		&	511.8	$\pm$ 	20.1		&	478.4	$\pm$ 	6.9		&	515.8	$\pm$ 	37.9		&	641.1	$\pm$ 	29.3	$\dagger$	&	513.7	$\pm$ 	18.1		&	523.6	$\pm$ 	35.5		&	495.4	$\pm$ 	39.9		\\
\OIII	&	4363.21	&	0.149	&	54.6	$\pm$ 	6.9		&	-		 		&	4.3	$\pm$ 	0.6		&	7.0	$\pm$ 	0.5		&	-		 		&	-		 		&	22.8	$\pm$ 	2.0		&	-		 		&	-		 		\\
\HeI	&	4387.93	&	0.141	&	-		 		&	-		 		&	4.6	$\pm$ 	0.5	$\dagger$	&	5.0	$\pm$ 	0.2		&	-		 		&	-		 		&	-		 		&	-		 		&	-		 		\\
\HeI	&	4471.47	&	0.115	&	-		 		&	40.3	$\pm$ 	3.4		&	38.6	$\pm$ 	1.5		&	36.9	$\pm$ 	0.7		&	-		 		&	-		 		&	45.1	$\pm$ 	3.0		&	-		 		&	-		 		\\
\FeIII	&	4658.10	&	0.058	&	-		 		&	-		 		&	5.9	$\pm$ 	0.6		&	2.1	$\pm$ 	0.3		&	-		 		&	-		 		&	-		 		&	-		 		&	-		 		\\
\ArIV+\HeI	&	4712.25	&	0.042	&	-		 		&	-		 		&	4.8	$\pm$ 	0.3		&	4.4	$\pm$ 	0.3		&	-		 		&	-		 		&	-		 		&	-		 		&	-		 		\\
\Hb	&	4861.33	&	0.000	&	1000.0	$\pm$ 	10.5		&	1000.0	$\pm$ 	5.6		&	1000.0	$\pm$ 	0.9		&	1000.0	$\pm$ 	5.4		&	1000.0	$\pm$ 	10.8		&	1000.0	$\pm$ 	24.7		&	1000.0	$\pm$ 	5.2		&	1000.0	$\pm$ 	11.2		&	1000.0	$\pm$ 	20.9		\\
\HeI	&	4921.93	&	-0.016	&	-		 		&	10.8	$\pm$ 	1.1		&	10.8	$\pm$ 	0.7		&	11.7	$\pm$ 	0.6		&	-		 		&	-		 		&	13.7	$\pm$ 	2.2	$\dagger$ 	&	-		 		&	-		 		\\
\OIII	&	4958.91	&	-0.026	&	1829.7	$\pm$ 	30.9		&	73.4	$\pm$ 	2.5		&	308.2	$\pm$ 	2.7		&	453.9	$\pm$ 	1.7		&	101.8	$\pm$ 	5.7		&	-		 		&	661.5	$\pm$ 	3.8		&	143.1	$\pm$ 	5.3		&	-		 		\\
\OIII	&	5006.84	&	-0.038	&	5804.3	$\pm$ 	55.0		&	235.3	$\pm$ 	4.2		&	937.2	$\pm$ 	9.0		&	1384.4	$\pm$ 	4.9		&	307.4	$\pm$ 	6.2		&	-		 		&	1997.0	$\pm$ 	16.7		&	416.2	$\pm$ 	8.0		&	-		 		\\
\HeI	&	5015.68	&	-0.040	&	-		 		&	16.2	$\pm$ 	2.3		&	24.5	$\pm$ 	1.0		&	27.0	$\pm$ 	0.3		&	-		 		&	-		 		&	35.5	$\pm$ 	4.0		&	-		 		&	-		 		\\
\ClIII	&	5517.71	&	-0.145	&	-		 		&	-		 		&	4.3	$\pm$ 	0.3		&	5.8	$\pm$ 	0.2		&	-		 		&	-		 		&	-		 		&	-		 		&	-		 		\\
\ClIII	&	5537.88	&	-0.149	&	-		 		&	-		 		&	4.6	$\pm$ 	0.3		&	4.7	$\pm$ 	0.3		&	-		 		&	-		 		&	-		 		&	-		 		&	-		 		\\
\NII	&	5754.64	&	-0.185	&	-		 		&	14.2	$\pm$ 	1.9		&	9.7	$\pm$ 	1.1		&	8.4	$\pm$ 	0.7		&	-		 		&	-		 		&	5.9	$\pm$ 	0.6		&	-		 		&	-		 		\\
\HeI	&	5875.64	&	-0.203	&	141.0	$\pm$ 	5.4		&	131.8	$\pm$ 	4.9	$\dagger$	&	123.6	$\pm$ 	3.1		&	116.9	$\pm$ 	1.2		&	-		 		&	-		 		&	141.6	$\pm$ 	7.0	$\dagger$ 	&	-		 		&	-		 		\\
\OI	&	6300.30	&	-0.263	&	-		 		&	-		 		&	4.0	$\pm$ 	0.4		&	14.4	$\pm$ 	0.2		&	-		 		&	-		 		&	-		 		&	-		 		&	-		 		\\
\SIII	&	6312.10	&	-0.264	&	15.7	$\pm$ 	0.9		&	7.5	$\pm$ 	1.0		&	14.4	$\pm$ 	0.5		&	16.2	$\pm$ 	0.5		&	-		 		&	-		 		&	16.1	$\pm$ 	1.4		&	-		 		&	-		 		\\
\SiII 	&	6347.11	&	-0.269	&	-		 		&	-		 		&	1.8	$\pm$ 	0.2		&	2.1	$\pm$ 	0.1		&	-		 		&	-		 		&	-		 		&	-		 		&	-		 		\\
\OI	&	6363.78	&	-0.271	&	-		 		&	10.6	$\pm$ 	0.5	$\dagger$	&	-		 		&	5.1	$\pm$ 	0.1		&	-		 		&	-		 		&	-		 		&	-		 		&	-		 		\\
\SiII 	&	6371.36	&	-0.272	&	-		 		&	-		 		&	-		 		&	1.2	$\pm$ 	0.2		&	-		 		&	-		 		&	-		 		&	-		 		&	-		 		\\
\NII	&	6548.03	&	-0.296	&	37.7	$\pm$ 	2.5		&	383.8	$\pm$ 	2.6		&	250.6	$\pm$ 	0.2		&	230.7	$\pm$ 	0.2		&	255.2	$\pm$ 	4.1		&	290.6	$\pm$ 	9.9		&	110.8	$\pm$ 	5.3		&	323.7	$\pm$ 	6.3		&	268.2	$\pm$ 	5.1		\\
\Ha	&	6562.82	&	-0.298	&	2820.0	$\pm$ 	2.2		&	2860.0	$\pm$ 	2.8		&	2860.0	$\pm$ 	4.1		&	2860.0	$\pm$ 	2.6		&	2860.0	$\pm$ 	5.3		&	2860.0	$\pm$ 	11.4		&	2820.0	$\pm$ 	2.5		&	2860.0	$\pm$ 	5.2		&	2860.0	$\pm$ 	7.2		\\
\NII	&	6583.41	&	-0.300	&	111.0	$\pm$ 	1.6		&	1157.5	$\pm$ 	2.3		&	771.4	$\pm$ 	0.9		&	712.3	$\pm$ 	0.7		&	745.1	$\pm$ 	7.2		&	919.5	$\pm$ 	12.1		&	335.0	$\pm$ 	3.9		&	966.7	$\pm$ 	5.7		&	813.8	$\pm$ 	8.4		\\
\HeI	&	6678.15	&	-0.313	&	38.1	$\pm$ 	0.8		&	28.0	$\pm$ 	0.5		&	30.4	$\pm$ 	0.4		&	29.4	$\pm$ 	0.2		&	-		 		&	-		 		&	33.0	$\pm$ 	0.5		&	34.4	$\pm$ 	2.3		&	-		 		\\
\SII	&	6716.47	&	-0.318	&	36.8	$\pm$ 	0.8		&	394.1	$\pm$ 	2.5		&	97.7	$\pm$ 	0.6		&	125.2	$\pm$ 	0.5		&	213.4	$\pm$ 	6.1		&	502.1	$\pm$ 	4.6		&	104.4	$\pm$ 	1.6		&	288.9	$\pm$ 	2.9		&	357.4	$\pm$ 	7.8		\\
\SII	&	6730.85	&	-0.320	&	31.0	$\pm$ 	0.9		&	344.2	$\pm$ 	2.4		&	115.9	$\pm$ 	0.7		&	159.5	$\pm$ 	0.6		&	164.0	$\pm$ 	7.2		&	382.3	$\pm$ 	5.0		&	85.1	$\pm$ 	1.2		&	240.4	$\pm$ 	3.3		&	350.7	$\pm$ 	7.8		\\
\HeI	&	7065.28	&	-0.365	&	45.7	$\pm$ 	0.7		&	18.9	$\pm$ 	0.7		&	37.4	$\pm$ 	0.7		&	24.3	$\pm$ 	0.2		&	23.6	$\pm$ 	1.1	$\dagger$	&	-		 		&	19.4	$\pm$ 	0.5		&	21.0	$\pm$ 	1.3		&	-		 		\\
\ArIII	&	7135.78	&	-0.374	&	116.7	$\pm$ 	2.8		&	79.2	$\pm$ 	1.9		&	119.4	$\pm$ 	2.3		&	102.7	$\pm$ 	0.7		&	72.4	$\pm$ 	2.0		&	-		 		&	118.6	$\pm$ 	2.0		&	107.1	$\pm$ 	2.6		&	-		 		\\
\CII	&	7231.34	&	-0.387	&	-		 		&	-		 		&	3.1	$\pm$ 	0.4	$\dagger$	&	-		 		&	-		 		&	-		 		&	-		 		&	-		 		&	-		 		\\
\HeI	&	7281.35	&	-0.393	&	-		 		&	-		 		&	6.2	$\pm$ 	0.4		&	-		 		&	-		 		&	-		 		&	-		 		&	-		 		&	-		 		\\
\OII	&	7319.46	&	-0.399	&	26.1	$\pm$ 	1.4	$\dagger$ 	&	38.3	$\pm$ 	3.3	$\dagger$	&	45.6	$\pm$ 	1.2	$\dagger$	&	42.5	$\pm$ 	0.5		&	66.3	$\pm$ 	5.4	$\dagger$	&	124.9	$\pm$ 	7.1	$\dagger$	&	27.6	$\pm$ 	1.1	$\dagger$ 	&	-		 		&	-		 		\\
\OII	&	7330.21	&	-0.400	&	14.5	$\pm$ 	1.5	$\dagger$ 	&	31.5	$\pm$ 	3.6	$\dagger$	&	37.4	$\pm$ 	1.0	$\dagger$	&	35.3	$\pm$ 	0.7		&	50.4	$\pm$ 	4.3	$\dagger$	&	79.8	$\pm$ 	6.6	$\dagger$	&	20.0	$\pm$ 	1.0	$\dagger$ 	&	-		 		&	-		 		\\
\ArIII	&	7751.10	&	-0.455	&	33.5	$\pm$ 	1.5		&	29.8	$\pm$ 	2.3		&	25.3	$\pm$ 	1.3		&	27.6	$\pm$ 	0.6		&	-		 		&	-		 		&	32.3	$\pm$ 	1.9		&	-		 		&	-		 		\\
P25	&	8323.42	&	-0.525	&	-		 		&	-		 		&	1.0	$\pm$ 	0.1		&	-		 		&	-		 		&	-		 		&	-		 		&	-		 		&	-		 		\\
P24	&	8333.78	&	-0.526	&	-		 		&	-		 		&	1.5	$\pm$ 	0.1		&	-		 		&	-		 		&	-		 		&	-		 		&	-		 		&	-		 		\\
P21	&	8374.48	&	-0.531	&	-		 		&	-		 		&	2.2	$\pm$ 	0.1		&	-		 		&	-		 		&	-		 		&	-		 		&	-		 		&	-		 		\\
P20	&	8392.40	&	-0.533	&	-		 		&	-		 		&	2.4	$\pm$ 	0.2		&	-		 		&	-		 		&	-		 		&	-		 		&	-		 		&	-		 		\\
P19	&	8413.20	&	-0.535	&	-		 		&	-		 		&	3.2	$\pm$ 	0.1		&	-		 		&	-		 		&	-		 		&	-		 		&	-		 		&	-		 		\\
P18	&	8437.95	&	-0.538	&	-		 		&	-		 		&	3.8	$\pm$ 	0.1		&	-		 		&	-		 		&	-		 		&	-		 		&	-		 		&	-		 		\\
\OI	&	8446.00	&	-0.538	&	-		 		&	-		 		&	4.1	$\pm$ 	0.1		&	-		 		&	-		 		&	-		 		&	-		 		&	-		 		&	-		 		\\
P17	&	8467.26	&	-0.541	&	-		 		&	-		 		&	4.8	$\pm$ 	0.1		&	-		 		&	-		 		&	-		 		&	-		 		&	-		 		&	-		 		\\
P16	&	8502.48	&	-0.544	&	-		 		&	-		 		&	5.3	$\pm$ 	0.1		&	-		 		&	-		 		&	-		 		&	-		 		&	-		 		&	-		 		\\
P15	&	8545.38	&	-0.549	&	-		 		&	-		 		&	6.1	$\pm$ 	0.1		&	7.3	$\pm$ 	0.2		&	-		 		&	-		 		&	-		 		&	-		 		&	-		 		\\
P14	&	8598.39	&	-0.554	&	7.8	$\pm$ 	0.7		&	-		 		&	7.4	$\pm$ 	0.1		&	7.7	$\pm$ 	0.2		&	-		 		&	-		 		&	6.7	$\pm$ 	0.5		&	-		 		&	-		 		\\
P12	&	8750.47	&	-0.568	&	-		 		&	-		 		&	11.3	$\pm$ 	0.5		&	-		 		&	-		 		&	-		 		&	-		 		&	-		 		&	-		 		\\
P11	&	8862.79	&	-0.578	&	-		 		&	-		 		&	15.8	$\pm$ 	0.5	$\dagger$	&	16.6	$\pm$ 	0.7		&	-		 		&	-		 		&	-		 		&	-		 		&	-		 		\\
P10	&	9015.30	&	-0.590	&	18.0	$\pm$ 	2.2		&	18.4	$\pm$ 	2.7		&	18.4	$\pm$ 	0.4		&	18.4	$\pm$ 	0.6		&	-		 		&	-		 		&	17.8	$\pm$ 	2.0	$\dagger$ 	&	18.4	$\pm$ 	1.2		&	-		 		\\
\SIII	&	9068.60	&	-0.594	&	212.2	$\pm$ 	5.3		&	198.0	$\pm$ 	4.0		&	370.8	$\pm$ 	1.1		&	420.8	$\pm$ 	0.8		&	211.2	$\pm$ 	14.5		&	105.8	$\pm$ 	7.1		&	201.4	$\pm$ 	12.8		&	145.9	$\pm$ 	1.2		&	73.5	$\pm$ 	6.1		\\
P9	&	9229.70	&	-0.604	&	32.8	$\pm$ 	1.1		&	32.1	$\pm$ 	0.7	$\dagger$ 	&	-		 		&	-		 		&	25.8	$\pm$ 	2.8		&	-		 		&	26.8	$\pm$ 	2.0		&	27.9	$\pm$ 	0.8		&	-		 		\\
\SIII$^{a}$ 	&	9530.60	&	-0.618	&	422.8	$\pm$ 	4.8		&	273.7	$\pm$ 	16.3		&	557.7	$\pm$ 	3.9		&	661.6	$\pm$ 	2.0		&	491.8	$\pm$ 	40.2		&	195.5	$\pm$ 	18.8		&	468.3	$\pm$ 	32.1		&	460.8	$\pm$ 	5.2	$\dagger$ 	&	192.1	$\pm$ 	12.1		\\
P8	&	9546.2	&	-0.618	&	44.4	$\pm$ 	3.3	$\dagger$ 	&	-		 		&	35.0	$\pm$ 	0.6		&	-		 		&	-		 		&	-		 		&	29.7	$\pm$ 	4.2		&	-		 		&	-		 		\\
P7	&	10049	&	-0.624	&	-		 		&	-		 		&	52.9	$\pm$ 	2.4		&	47.8	$\pm$ 	1.5		&	-		 		&	-		 		&	41.9	$\pm$ 	4.0	$\dagger$ 	&	-		 		&	-		 		\\
\\ \\ 																																									
F(\Hb)$^{b}$ 	&		&		&	5.41	$\pm$ 	1.30		&	2.63	$\pm$ 	0.78		&	39.49	$\pm$ 	9.93		&	17.55	$\pm$ 	1.61		&	23.61	$\pm$ 	5.25		&	7.73	$\pm$ 	1.12		&	68.26	$\pm$ 	14.59		&	61.83	$\pm$ 	15.67		&	27.98	$\pm$ 	3.74		\\
c(\Hb) 	&		&		&	2.66	$\pm$ 	0.10		&	1.36	$\pm$ 	0.13		&	1.57	$\pm$ 	0.11		&	0.98	$\pm$ 	0.04		&	1.50	$\pm$ 	0.10		&	1.41	$\pm$ 	0.06		&	1.42	$\pm$ 	0.09		&	1.54	$\pm$ 	0.11		&	1.91	$\pm$ 	0.06		\\
 \hline
\end{longtable}
\end{center}
     \begin{list}{}{} \footnotesize{
			\item NOTES:
			\item $\dagger$ Lines with uncertain measures. These lines are not used in the estimation of physical parameters or chemical abundances.
			\item (a)=Line strongly affected by a sky absorption band. Not used in analysis.
			\item (b)= Reddening-corrected \Hb~line fluxes in units of 10$^{-13}$~erg~cm$^{-2}$~s$^{-1}$.
			} 
		\end{list}
\end{landscape}
\normalsize
\twocolumn

\begin{landscape} 
\setcounter{table}{4} 
\begin{table}[h!]
\caption{Ionic and total chemical abundances for all \HII~regions of the sample with direct T$_e$ estimations.} 
\label{table:abundances}
 \setlength\tabcolsep{2pt}
\centering                
\begin{tabular}{l c c c c c c c c c c c c c}
\\
\hline
		&S83		&		S127			&	S128			&	S132			&	S156		&		S158		&			S162		&		S206		&		S212			&	S255		&		S298			&	S301 	&			S311 \\			
\hline \hline   \\
12+log(O$^+$/H$^+$)	&	7.30	$\pm$ 	0.06	&	8.36	$\pm$ 	0.12	&	8.06	$\pm$ 	0.24	&	8.30	$\pm$ 	0.11	&	8.23	$\pm$ 	0.09	&	8.10	$\pm$ 	0.30		&	8.25	$\pm$ 	0.07	&	8.17	$\pm$ 	0.12	&	7.88	$\pm$ 	0.10	&	8.87			&	7.82	$\pm$ 	0.11	&	8.08			&	8.07			\\
12+log(O$^{2+}$/H$^+$) 	&	8.09	$\pm$ 	0.06	&	7.59	$\pm$ 	0.19	&	8.10	$\pm$ 	0.26	&	7.21	$\pm$ 	0.15	&	7.62	$\pm$ 	0.05	&	8.14	$\pm$ 	0.40		&	7.75	$\pm$ 	0.03	&	8.11	$\pm$ 	0.03	&	7.56	$\pm$ 	0.04	&	7.47			&	8.22	$\pm$ 	0.03	&	7.80			&	8.19			\\
12+log(S$^+$/H$^+$) 	&	5.05	$\pm$ 	0.03	&	5.77	$\pm$ 	0.07	&	5.68	$\pm$ 	0.14	&	6.43	$\pm$ 	0.13	&	5.88	$\pm$ 	0.25	&	5.41	$\pm$ 	0.16		&	6.11	$\pm$ 	0.05	&	-	 		&	5.53	$\pm$ 	0.05	&	6.10			&	5.93	$\pm$ 	0.06	&	5.83			&	5.93			\\
12+log(S$^{2+}$/H$^+$)	&	6.33	$\pm$ 	0.02	&	6.69	$\pm$ 	0.18	&	6.89	$\pm$ 	0.26	&	6.60	$\pm$ 	0.03	&	6.89	$\pm$ 	0.01	&	7.39	$\pm$ 	0.88	$\dagger$	&	6.94	$\pm$ 	0.04	&	-	 		&	6.29	$\pm$ 	0.03	&	-			&	6.63	$\pm$ 	0.10	&	-			&	7.32			\\
12+log(N$^+$/H$^+$) 	&	6.18	$\pm$ 	0.03	&	7.11	$\pm$ 	0.08	&	6.89	$\pm$ 	0.15	&	7.40	$\pm$ 	0.06	&	7.22	$\pm$ 	0.05	&	6.94	$\pm$ 	0.17		&	7.23	$\pm$ 	0.04	&	-	 		&	6.70	$\pm$ 	0.06	&	7.45			&	6.78	$\pm$ 	0.07	&	7.03			&	6.94			\\
log(N$^+$/O$^+$) 	&	-1.09	$\pm$ 	0.04	&	-1.21	$\pm$ 	0.06	&	-1.16	$\pm$ 	0.11	&	-0.87	$\pm$ 	0.05	&	-0.91	$\pm$ 	0.04	&	-1.04	$\pm$ 	0.14		&	-0.90	$\pm$ 	0.03	&	-	 		&	-1.16	$\pm$ 	0.04	&	-1.39			&	-1.03	$\pm$ 	0.05	&	-1.04			&	-1.11			\\
12+log(Ne$^{2+}$/H$^+$) 	&	7.57	$\pm$ 	0.09	&	-	 		&	-	 		&	-	 		&	6.33	$\pm$ 	0.08	&	-	 			&	6.57	$\pm$ 	0.04	&	-	 		&	-	 		&	-			&	7.82	$\pm$ 	0.04	&	7.04			&	7.43			\\
12+log(Ar$^{2+}$/H$^+$) 	&	5.91	$\pm$ 	0.02	&	6.05	$\pm$ 	0.21	&	6.32	$\pm$ 	0.31	&	6.10	$\pm$ 	0.03	&	6.29	$\pm$ 	0.02	&	6.28	$\pm$ 	0.47	$\dagger$	&	6.23	$\pm$ 	0.05	&	-	 		&	5.88	$\pm$ 	0.02	&	-			&	6.09	$\pm$ 	0.04	&	6.08			&	6.39			\\
12+log(Fe$^{2+}$/H$^+$) 	&	-	 		&	-	 		&	-	 		&	-	 		&	5.70	$\pm$ 	0.07	&	-	 			&	5.21	$\pm$ 	0.07	&	-	 		&	-	 		&	-			&	-	 		&	-			&				\\
(He$^+$/H$^+$) 4026	&	-	 		&	-	 		&	-	 		&	-	 		&	0.08	$\pm$ 	0.01	&	-	 			&	0.09	$\pm$ 	0.01	&	-	 		&	0.05	$\pm$ 	0.01	&	-			&	-	 		&	-			&	0.09			\\
(He$^+$/H$^+$) 4471 	&	-	 		&	-	 		&	-	 		&	0.08	$\pm$ 	0.01	&	0.08	$\pm$ 	0.01	&	-	 			&	0.08	$\pm$ 	0.01	&	-	 		&	0.09	$\pm$ 	0.01	&	-			&	-	 		&	0.09			&	0.11			\\
(He$^+$/H$^+$) 5875 	&	0.10	$\pm$ 	0.01	&	0.07	$\pm$ 	0.01	&	0.06	$\pm$ 	0.01	&	-	 		&	0.09	$\pm$ 	0.01	&	0.10	$\pm$ 	0.02		&	0.09	$\pm$ 	0.01	&	0.09	$\pm$ 	0.01	&	-	 		&	-			&	0.09	$\pm$ 	0.01	&	0.09			&	-			\\
(He$^+$/H$^+$) 6678	&	0.10	$\pm$ 	0.01	&	0.07	$\pm$ 	0.01	&	0.10	$\pm$ 	0.02	&	0.07	$\pm$ 	0.01	&	0.08	$\pm$ 	0.01	&	-	 			&	0.08	$\pm$ 	0.01	&	-	 		&	0.09	$\pm$ 	0.01	&	-			&	0.10	$\pm$ 	0.02	&	0.08			&	0.10			\\
(He$^+$/H$^+$) 7065	&	0.11	$\pm$ 	0.01	&	0.05	$\pm$ 	0.01	&	0.08	$\pm$ 	0.02	&	0.07	$\pm$ 	0.01	&	0.09	$\pm$ 	0.01	&	-	 			&	0.06	$\pm$ 	0.01	&	-	 		&	0.05	$\pm$ 	0.01	&	-			&	-	 		&	0.08			&	0.08			\\
(He$^+$/H$^+$) 	&	0.10	$\pm$ 	0.01	&	0.06	$\pm$ 	0.01	&	0.07	$\pm$ 	0.01	&	0.07	$\pm$ 	0.01	&	0.08	$\pm$ 	0.01	&	0.10	$\pm$ 	0.02		&	0.08	$\pm$ 	0.01	&	0.09	$\pm$ 	0.01	&	0.07	$\pm$ 	0.01	&	-			&	0.09	$\pm$ 	0.01	&	0.09			&	0.09			\\
\\																																																						
ICF(S$^+$,S$^{2+}$)	&	1.60	$\pm$ 	0.09	&	1.00	$\pm$ 	0.01	&	1.09	$\pm$ 	0.10	&	1.00	$\pm$ 	0.01	&	1.01	$\pm$ 	0.01	&	1.09	$\pm$ 	0.15		&	1.01	$\pm$ 	0.01	&	-	 		&	1.02	$\pm$ 	0.01	&	-			&	1.26	$\pm$ 	0.07	&	-			&	1.12			\\
ICF(Ne$^{2+}$) 	&	1.07	$\pm$ 	0.01	&	-	 		&	1.15	$\pm$ 	0.09	&	-	 		&	1.65	$\pm$ 	0.16	&	-	 			&	1.49	$\pm$ 	0.09	&	-	 		&	-	 		&	-			&	1.09	$\pm$ 	0.01	&	1.30			&	1.13			\\
ICF(Ar$^{2+}$) 	&	1.28	$\pm$ 	0.07	&	1.26	$\pm$ 	0.03	&	1.12	$\pm$ 	0.04	&	1.29	$\pm$ 	0.01	&	1.24	$\pm$ 	0.02	&	1.13	$\pm$ 	0.06		&	1.22	$\pm$ 	0.01	&	-	 		&	1.19	$\pm$ 	0.02	&	-			&	1.12	$\pm$ 	0.02	&	1.18			&	1.12			\\
ICF(Fe$^{2+}$) 	&	-	 		&	-	 		&	-	 		&	-	 		&	1.41	$\pm$ 	0.04	&	-	 			&	1.46	$\pm$ 	0.04	&	-	 		&	-	 		&	-			&	-	 		&	-			&				\\
ICF(He$^+$) 	&	1.03	$\pm$ 	0.01	&	1.12	$\pm$ 	0.05	&	1.06	$\pm$ 	0.04	&	1.66	$\pm$ 	0.36	&	1.10	$\pm$ 	0.06	&	-				&	1.14	$\pm$ 	0.02	&	-			&	1.17	$\pm$ 	0.03	&	-			&	1.15	$\pm$ 	0.04	&	-			&	1.04			\\
\\																																																						
12+log(O/H) 	&	8.16	$\pm$ 	0.05	&	8.42	$\pm$ 	0.11	&	8.38	$\pm$ 	0.18	&	8.33	$\pm$ 	0.10	&	8.33	$\pm$ 	0.07	&	8.42	$\pm$ 	0.25		&	8.37	$\pm$ 	0.05	&	8.44	$\pm$ 	0.06	&	8.05	$\pm$ 	0.07	&	$<$8.89			&	8.37	$\pm$ 	0.04	&	$>$8.26			&	8.44	$\pm$ 	0.04	\\
12+log(S/H) 	&	6.56	$\pm$ 	0.03	&	6.74	$\pm$ 	0.16	&	6.96	$\pm$ 	0.24	&	6.83	$\pm$ 	0.06	&	6.93	$\pm$ 	0.03	&	7.43	$\pm$ 	0.88	$\dagger$	&	7.01	$\pm$ 	0.04	&	-	 		&	6.37	$\pm$ 	0.03	&	-			&	6.81	$\pm$ 	0.09	&	-			&	7.39	$\pm$ 	0.10	\\
12+log(N/H) 	&	7.04	$\pm$ 	0.08	&	7.18	$\pm$ 	0.18	&	7.21	$\pm$ 	0.33	&	7.44	$\pm$ 	0.16	&	7.32	$\pm$ 	0.12	&	7.26	$\pm$ 	0.43		&	7.35	$\pm$ 	0.09	&	-	 		&	6.87	$\pm$ 	0.13	&	$>$7.46			&	7.33	$\pm$ 	0.14	&	$>$7.22			&	7.31	$\pm$ 	0.14	\\
log(N/O) 	&	-1.09	$\pm$ 	0.04	&	-1.21	$\pm$ 	0.06	&	-1.16	$\pm$ 	0.11	&	-0.87	$\pm$ 	0.05	&	-0.91	$\pm$ 	0.04	&	-1.04	$\pm$ 	0.14		&	-0.90	$\pm$ 	0.03	&	-	 		&	-1.16	$\pm$ 	0.04	&	$>$-1.39			&	-1.03	$\pm$ 	0.05	&	$<$-1.04			&	-1.11	$\pm$ 	0.06	\\
12+log(Ne/H) 	&	7.60	$\pm$ 	0.09	&	-	 		&	-	 		&	-	 		&	6.55	$\pm$ 	0.09	&	-	 			&	6.74	$\pm$ 	0.05	&	-	 		&	-	 		&	-			&	7.86	$\pm$ 	0.04	&	$>$7.15			&	7.49	$\pm$ 	0.04	\\
12+log(Ar/H) 	&	6.02	$\pm$ 	0.03	&	6.15	$\pm$ 	0.21	&	6.37	$\pm$ 	0.31	&	6.21	$\pm$ 	0.03	&	6.39	$\pm$ 	0.02	&	6.33	$\pm$ 	0.47	$\dagger$	&	6.31	$\pm$ 	0.05	&	-	 		&	5.96	$\pm$ 	0.03	&	-			&	6.14	$\pm$ 	0.04	&	$>$6.15			&	6.43	$\pm$ 	0.04	\\
12+log(Fe/H) 	&	-	 		&	-	 		&	-	 		&	-	 		&	5.85	$\pm$ 	0.07	&	-	 			&	5.38	$\pm$ 	0.07	&	-	 		&	-	 		&	-			&	-	 		&	-			&	-			\\
He/H 	&	0.10	$\pm$ 	0.01	&	0.07	$\pm$ 	0.01	&	0.07	$\pm$ 	0.01	&	0.12	$\pm$ 	0.03	&	0.09	$\pm$ 	0.01	&	-				&	0.09	$\pm$ 	0.01	&	-			&	0.08	$\pm$ 	0.01	&	-			&	0.10	$\pm$ 	0.01	&	-			&	0.09	$\pm$ 	0.01	\\
\\
\hline
\end{tabular}
     \begin{list}{}{} \footnotesize{
			\item $\dagger$ Uncertain abundances derived from T$_e$(\SIII) with S/N$<$3.} 
		\end{list}
		\end{table}
\end{landscape}

\end{document}